\documentclass[english,aps,10pt,pre,a4paper, floatfix, notitlepage, nofootinbib, reprint]{revtex4-1}

\newif\iftwocol
\twocoltrue
\usepackage[normalem]{ulem}

\usepackage{lmodern}
\usepackage{amsmath}
\usepackage{amsfonts}
\usepackage{dsfont}
\usepackage{graphicx}
\usepackage{xcolor}
\usepackage{braket}
\usepackage[format=plain,justification=centerlast]{caption}
\usepackage{subcaption}

\usepackage{mathtools}

\usepackage[version=4]{mhchem}

\usepackage{tikz}

\usetikzlibrary{arrows.meta}
\usetikzlibrary{patterns}
\usepackage[compat=1.1.0]{tikz-feynman}

\usepackage[export]{adjustbox}

\usepackage[unicode=true,pdfusetitle,
bookmarks=true,bookmarksnumbered=true,bookmarksopen=true,bookmarksopenlevel=2,
breaklinks=false,pdfborder={0 0 0},backref=false,colorlinks=true]{hyperref}
\usepackage{bookmark}

\DeclareMathOperator{\tr}{\mathrm{tr}}
\renewcommand{\d}{\dagger}
\newcommand{\bubblesize}{1.5}

\iftwocol
\newcommand*{\tcbr}{\nonumber \\ &}
\else
\newcommand*{\tcbr}{}
\fi

\newcommand*{\TNS}{\ce{Ta2NiSe5}}

\newcommand*{\heading}[1]{\belowpdfbookmark{#1}{#1}{\bfseries\textit{#1.---}}\ignorespaces}

\renewcommand*{\revised}[1]{#1}

\newcommand*{\del}{\nabla}

\newlength{\apsfigurewidth}
\usepackage{comment}
\iffalse

\else
\excludecomment{gobble}
\fi

\def \bubble {\includegraphics[raise=-3.5mm]{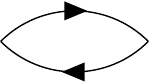}}

\def \bubbleRPA {\includegraphics[raise=-3.5mm]{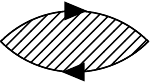}}

\def \bubbleDisordered {\includegraphics[raise=-3.5mm]{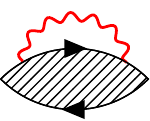}}

\def \bubbleRPADisordered{\includegraphics[raise=-3.5mm]{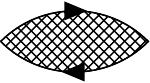}}

\def \bubblePurse {
	\begin{tikzpicture}[baseline={([yshift=-.5ex]m.base)}]
	\begin{feynman}
	\vertex (m) at (0,0) ;
	\vertex (n) at (+\bubblesize,0) ;
	\vertex (u1) at (\bubblesize*0.5 - \bubblesize*0.353553, \bubblesize*0.414213*0.5 - \bubblesize*0.0947343);
	\vertex (u2) at (\bubblesize*0.5 + \bubblesize*0.353553, \bubblesize*0.414213*0.5 - \bubblesize*0.0947343);
	
	\diagram* {
		(m) -- [out = 45, in = 135] (n) -- [out = -135, in = -45] (m),
		(u1) -- [boson, red, thick, half left] (u2),
	};
	\end{feynman}
	\end{tikzpicture}
}

\def \bubbleClam {
	\begin{tikzpicture}[baseline={([yshift=-.5ex]m.base)}]
	\begin{feynman}
	\vertex (m) at (0,0) ;
	\vertex (n) at (+\bubblesize,0) ;
	\vertex (u1) at (\bubblesize*0.5,  \bubblesize*0.414213*0.5);
	\vertex (u2) at (\bubblesize*0.5, -\bubblesize*0.414213*0.5);
	
	\diagram* {
		(m) -- [out = 45, in = 135] (n) -- [out = -135, in = -45] (m),
		(u1) -- [boson, red, thick] (u2),
	};
	\end{feynman}
	\end{tikzpicture}
}

\def \disorderprop {
	\feynmandiagram [small, baseline={([yshift=-0.5ex]current bounding box.center)}, horizontal=a to b] {
		a -- [red, boson] b
	};}

\begin{document}

	\title{\Large Effects of Disorder on the Transport of Collective Modes in an Excitonic Condensate}
	
	\author{Benjamin Remez}
	\affiliation{T.C.M. Group, Cavendish Laboratory, University of Cambridge, JJ Thomson Avenue,	Cambridge CB3 0HE, United Kingdom}
	\author{Nigel R. Cooper}
	\affiliation{T.C.M. Group, Cavendish Laboratory, University of Cambridge, JJ Thomson Avenue,	Cambridge CB3 0HE, United Kingdom}

	\begin{abstract}
		An excitonic insulator (EI) is an unconventional quantum phase of matter in which excitons, bound pairs of electrons and holes, undergo Bose--Einstein condensation, forming a macroscopic coherent state. While its existence was first hypothesized half a century ago, the EI has eluded experimental observation in bulk materials for decades. In the last few years, a resurgence of interest in the subject has been driven by the identification of several candidate materials suspected to support an excitonic condensate. However, one obstacle in verifying the nature of these systems has been to find signatures of the EI that distinguish it from a normal insulator. To address this, we focus on a clear qualitative difference between the two phases: the existence of Goldstone modes born by the spontaneous breaking of a $U(1)$ symmetry in the EI. Even if this mode is gapped, as occurs in the case of an approximate symmetry, this branch of collective modes remains a fundamental feature of the low-energy dynamics of the EI provided the symmetry-breaking is small. We study a simple model that realizes an excitonic condensate, and use mean field theory within the random-phase approximation to determine its collective modes. We subsequently develop a diagrammatic method to incorporate the effects of disorder perturbatively, and use it to determine the scattering rate of the collective modes. We interpret our results within an effective field theory. The collective modes are found to be robust against symmetry-preserving disorder, implying an experimental fingerprint unique to the EI: the ballistic propagation of low-lying modes over mesoscopic distances, at electronic-scale velocities. We suggest this could affect thermal transport at low temperatures, and could be observed via spatially-resolved pump-probe spectroscopy through the coherent response of phonons that hybridize with the collective modes.
	\end{abstract}
	\date{\today}
	
	\maketitle
	
	%\tableofcontents
	%\makeatletter
	%\let\toc@pre\relax
	%\let\toc@post\relax
	%\makeatother
	
	%\pagebreak

	\section{Introduction}
	
	The excitonic insulator (EI) is an unconventional phase of matter in which excitons condense into a macroscopically phase-coherent state \cite{Jerome1967,Kohn1967,Kohn1967mbp,Halperin1968}. 
	This phase lies at the transition between semiconductors and semimetals \cite{Mott1968, Imada1998, Bronold2006}, and can be understood in terms of Bose-Einstein condensation (BEC) \cite{BECPitaevskiiStringari} or the Bardeen--Cooper--Schrieffer (BCS) \cite{BCS1957} theory of superconductivity \cite{Bronold2006,Phan2010,Seki2011,Perfetto2019}.
	The EI is expected to exist in equilibrium and possibly at room temperature, and thus provide unprecedented access for the study of the BCS-BEC crossover, the metal-insulator transition and other strongly-correlated electron systems, as well as  potential applications such as thermoelectrics \cite{Wu2014} and dynamical band engineering \cite{Mor2017, Okazaki2018}. 
	
	Although it was first hypothesized decades ago, recent years have seen renewed interest in the EI, with the theoretical and experimental identification of several candidate materials, including transition metal \revised{chalcogenides} such as \TNS{}  \cite{Wakisaka2009, Kaneko2012, Kaneko2013,Seki2014, Kim2016,Sugimoto2016,Sugimoto2016a,Lu2017,Larkin2017, Mor2017, Larkin2018,Seo2018, Werdehausen2018ScienceAdvances,Okazaki2018, Domon2018,Lee2019}, \emph{1T}-\ce{TiSe2} \cite{DiSalvo1976,Pillo2000, Cercellier2007, Monney2009,Rossnagel2011,Hellmann2012, Kogar2017} and $\ce{TmSe_{1-x}Te_x}$ \cite{Wachter2004,Hulsen2006,Wachter2013}, as well as low-dimensional systems such as graphene, quantum wells, and quantum Hall bilayers \cite{Eisenstein2014,Zhu1995,Naveh1996,Khveshchenko2001, Eisenstein2004,Aleiner2007,Wu2015,Du2017,Varsano2017}.
	
	Despite this influx of candidate materials, confirmation of the existence of the EI phase in bulk materials has remained elusive. This is due to measured EI signatures being either hard to quantify, or difficult to separate from those of a normal insulator \cite{counterflow}. This situation signals an urgent need for a general distinguishing experimental signature which is unique to the EI.

	To address this issue, we argue that a key distinguishing feature of the ideal EI is that it exhibits a spontaneously broken symmetry. 
	The symmetry in question is {$U(1)$}, corresponding to the independent conservation of charge in the conduction and valence bands \cite{Phan2010,Wu2015}. While in practice this symmetry is not exact and different bands are generally coupled, it nevertheless can be approximately conserved if tunnelling between the bands is heavily suppressed. This could be due to a myriad of causes: the bands existing in two spatially separated planes (as in bilayer and quantum-well systems \cite{Eisenstein2004}); the localization of band orbitals around different atoms in the unit cell of \TNS{} \cite{Kaneko2013}; and the indirectness of the band gap in \emph{1T}-\ce{TiSe2} \cite{Rossnagel2011,Kogar2017}. Upon formation of the condensate, this symmetry spontaneously breaks. {(There is a separate $U(1)$ symmetry reflecting total charge conservation, which remains unbroken.)} 
	Breaking of such a continuous symmetry gives rise to gapless Nambu--Goldstone collective modes (CMs) \cite{Nambu1960, Goldstone1961, Goldstone1962}. 
	If the symmetry is not exact, the CMs will be gapped. Strong symmetry breaking could push the CM gap above the particle-hole continuum. However, if the symmetry violation is small, then the CM branch will remain low-lying. Therefore we suggest that the observation of soft modes, below the particle-hole continuum, is characteristic of an EI, providing vestigial evidence of the breaking of an underlying (approximate) symmetry.
	Since such collective modes can be taken as evidence for an underlying EI phase, we 
	study the properties of these modes.  We focus on their robustness to disorder scattering, and outline their experimental consequences.

	We consider a minimal model which exhibits an excitonic phase transition: \revised{an extended Falicov--Kimball model \cite{Falicov1969} is driven into the EI phase by strong on-site electron--electron repulsion \cite{Batista2002, Batista2004, Ejima2014, Golez2016, Murakami2017,Tanabe2018, Murakami2020}. } The CMs are found via the density-density response function  $\chi\left(k,\omega\right)$ (defined below) within the random-phase approximation (RPA) \cite{Bohm1953, Pines1953, Chen2017}. We subsequently demonstrate the robustness of the CMs by calculating their mean free paths in the presence of a  disorder potential. Disorder can couple to the excitonic insulator in a variety of different channels, which we describe. \revised{In this we generalize Zittartz's previous study \cite{Zittartz1967} of EI quasiparticle disorder scattering and extend it to the collective modes.} We show that at small wavevectors the excitonic modes can have mean free paths orders of magnitude longer than those of single quasiparticles. We show that this can be  attributed to the relation between the CMs and the condensate broken symmetry, as well as their non-zero group velocity at small $k$.
	The large CM mean free paths suggest they should have an appreciable signature in thermal transport, and could potentially be observed directly in spatially-resolved ultrafast pump-probe experiments.

	The rest of this paper is organized as follows: In Sec.~\ref{sec:Model} we  present our model and the formalism of our diagrammatic linear response treatment, and we subsequently employ the RPA to identify the collective modes. We extend the formalism to incorporate disorder in Sec.~\ref{sec:Disorder}, and compute the disorder-induced scattering rates of both single particles and collective modes. In Sec.~\ref{sec:Discussion} we discuss our results qualitatively, formulate an effective field theory whereby we can interpret all of the results in a compact general fashion, and make connections to possible experimental consequences. Technical details of our calculations are given in the Appendices.

	\section{Excitonic Condensation in The Extended Falicov--Kimball Model} \label{sec:Model}

	\begin{figure}[tb]
		\centering
		\includegraphics[width=\apsfigurewidth]{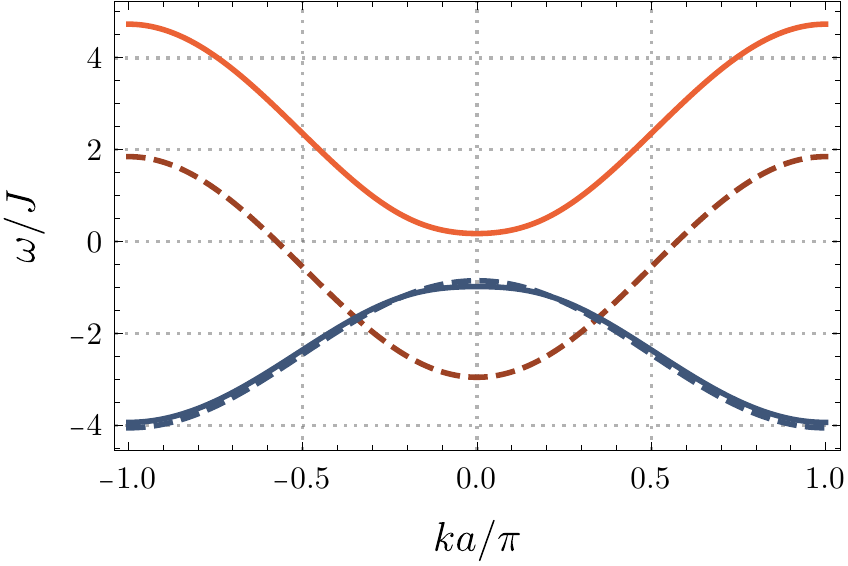}
		\caption{The non-interacting (dashed) and interacting (solid) electron bands. \revised{The interaction separates the overlapping bare bands into a semiconducting configuration. The subsequent condensation of excitons} 
		%The interaction drives the bands from a semimetallic configuration to an excitonic insulating one. The excitonic nature of the new state 
		is glimpsed in the modest flattening of the valence band, indicating the hybridization of the original bands --- i.e. a correlated coexistence of electrons with holes.}
		\label{fig:quasiparticlebands}
	\end{figure}
	
	\heading{Model}
	We consider a simple model that exhibits excitonic condensation \revised{\cite{Batista2002, Batista2004, Ejima2014, Golez2016, Murakami2017,Tanabe2018, Murakami2020}.  The extended Falicov--Kimball model consists of itinerant spinless conduction and valence electrons, with a repulsive interaction between colocated particles:} 
	%It consists of spinless electrons in a one-dimensional lattice, with repulsive interactions between electrons in conduction and valence states:
	\begin{equation} \label{eq:Hubbard_Hamiltonian}
	H = \!\!\sum_{k,\,\alpha=c,v}\!\!\left(\epsilon_\alpha(k)+\Delta_\alpha \right)\tilde{c}_{k,\alpha}^\d \tilde{c}^{\phantom{\dagger}}_{k,\alpha}
	+U_0 \sum_{i=1}^{N} n_{i,c} n_{i,v}\,.
	\end{equation}
	$\tilde{c}_{k,c \left(v\right)}  = \frac{1}{\sqrt{N}} \sum_m e^{-ikm} c_{m,c \left(v\right)}$ are the annihilation operators in the conduction (valence) band, $\epsilon_{\alpha}(k)$ the band dispersions and $\Delta_\alpha$ the band offsets. $U_0 > 0$ is the on-site repulsion which drives the formation of the condensate. $N$ is the number of sites.
	We set $\hbar = 1$ throughout. The lattice constant is $a=1$, so that the first Brillouin zone (BZ) is $k\in\left[-\pi,\pi\right]$.
	We use nearest-neighbour hopping bands $\epsilon_{c,v}\left(k\right) = \mp J_{c,v} \cos {k}$, and our unit of energy is the average transfer integral $J = \frac{1}{2}\left(J_c + J_v \right)$.
	
	\revised{Ultimately we will be interested in bulk 3d systems, where long-range excitonic order is possible. This allows us to rely on mean-field theory, apply the RPA, and neglect localization due to disorder, assuming it is weak. While the numerical results below are calculated in one dimension for practicality, the formalism is readily evaluated in higher dimensions and will produce qualitatively similar results. We will point out  differences due to dimensionality where they arise. From these one-dimensional results we will infer the properties of 3d systems. This will be supported by an effective field theory which we solve in three dimensions.}

	$H$ has a manifest $U(1)$ symmetry, as it does not mix states in different bands. In the excitonic insulator phase this symmetry will be broken, and the ground state of the system will have a non-zero value for the complex-valued order parameter 
	\begin{equation}
	\phi_i =\braket{c^\d_{i,c} c^{\phantom{\dagger}}_{i,v}} \neq 0 \,.
	\end{equation}
	
	We determine the order parameter for this condensed ground state within the mean-field approximation, using standard methods as described in Appendix \ref{app:GroundStateProperties}.
	As a starting point for our model parameters we choose $\Delta_c = -0.55~J,\, \Delta_v = -2.45~J,$ $U = 3.0~J$ and $J_c=J_v=J$. These values \revised{produce a semiconducting normal phase}, match \cite{Murakami2017} angle-resolved photoelectron spectroscopy measurements performed on \TNS{} \cite{Mor2017}, and place the model in the BEC regime \cite{Phan2010} (see Appendix \ref{app:GroundStateProperties}). However, we modify the transfer integrals to explicitly break particle--hole symmetry, setting $J_v = 1.2~J$ and $J_c = 0.8~J$. This is because electron--hole symmetry will protect the CMs from certain disorder scattering mechanisms, but in general we do not expect such a fine-tuned symmetry. \revised{The bare and interaction-driven quasiparticle bands, $\epsilon_\alpha (k)$ and $E_\pm(k)$, respectively,} are depicted in Fig.~\ref{fig:quasiparticlebands}. The system is at half-filling if the chemical potential is fixed at $\mu=0$, and we henceforth suppress it.

	\heading{Linear response}
	We investigate the collective modes through their signatures in the system linear response \cite{Murakami2020, Murakami2020Phonons}. 
	The observables whose response we are interested in are the electron density operators. We define them in real space as the local spinor vectors
	\begin{equation}
	\hat{n}^\mu_i = \Psi_i^\d \sigma^\mu \Psi_i^{\phantom{\dagger}},\quad (\mu=0,\,x,\,y,\,z)
	\end{equation}
	with $ \sigma^0 = \mathds{1}$ the identity matrix and the rest Pauli's, and defining the vector of annihilation operators $\Psi_i~=~(c_{i,c},~c_{i,v})^{\rm T}$, with the corresponding Fourier transform $\tilde{\Psi}_k$. The mean fields correspond to the expectation values of $\hat{n}^\mu$, as $\braket{\phi} = \frac{1}{2} \braket{\hat{n}^x + i \hat{n}^y }$ and $\braket{n_{c,v}} = \frac{1}{2} \braket{\hat{n}^0 \pm \hat{n}^z}$. $\hat{n}^0$ is the total electron density operator.
	
	The susceptibility is given by the Kubo formula \cite[Ch. 7]{AltlandSimons2010}
	\begin{align} 
	\chi_{\mu\nu}\left(i,t\right) 
	& = -i\Theta\left(t\right) \frac{\braket{GS|\left[ \hat{n}_i^\mu\left(t\right) , \hat{n}_0^\nu \left(0\right) \right]|GS}}{\braket{GS|GS}}\,, \label{ChiWickTheoremNew} 
	\end{align}
	evaluated with respect to the (interacting) ground state (GS) of the system, yielding the retarded response function. 
	This susceptibility is related to the time-ordered response function, which admits a diagrammatic expansion via Wick's Theorem \cite{PeskinSchroeder1995}. This is written as \cite{FetterWalecka}
	\begin{align} \label{eq:ChiWickTheoremConnected} 
	\chi_{\mu\nu}\left(i,t\right) 
	& = -i \Braket{GS\left|\mathcal{T} \lbrace \hat{n}_i^\mu\left(t\right) \hat{n}_0^\nu \left(0\right) \rbrace \right|GS}_\text{conn.}^\text{R}
	\tcbr = -i \times  \,
	\left[\sigma^\mu
	\begin{tikzpicture}[baseline={([yshift=-0.5ex]current bounding box.center)}]
	\begin{feynman}
	\vertex (m) at (-1.5,0) ;
	\vertex (n) at (+1.5,0) ;
	\vertex (ru) at (+0.5, +0.5) ;
	\vertex (rd) at (+0.5, -0.5) ;
	\vertex (lu) at (-0.5, +0.5) ;
	\vertex (ld) at (-0.5, -0.5) ;
	\diagram* {
		(m) -- [fermion, bend left = 20] (lu) -- (ru) -- [fermion, bend left = 20] (n),
		(n) -- [fermion, bend left = 20] (rd) -- (ld) -- [fermion, bend left = 20] (m),
	};
	\draw [fill=gray] (ld) rectangle (ru);
	\end{feynman}
	\end{tikzpicture} \, \sigma^\nu \right]^\text{R} \,.
	\end{align}
	Here $\mathcal{T}$ is the time-ordering operator. The notation $\left[ \dots \right]^\text{R}$ signifies that the linear response function $\chi$ is obtained by retarding the quantity in brackets, i.e. pushing all its poles to the lower half of the complex frequency plane \cite[p. 174]{FetterWalecka}. Henceforth we carry out calculations in terms of time-ordered functions, and the retardation at the end of the procedure is implied. The diagrammatic expansion for $\chi$ is the sum of all connected diagrams with two external vertices, as shown.
	
	The collective modes are most easily identified in $k$-space. 
	Assuming translational invariance, one obtains  
	\begin{equation} \label{eq:Chi0KuboMomentumFrequencySpace}
	\chi_{\mu\nu} \! \left(k,\omega\right) 
	= \left[ \frac{1}{\omega+i\eta} \right]\!\! * \!\!
	\int_{-\infty}^{\infty}\!\!\!\!\!
	\Braket{\left[  \hat{n}_{k}^\mu\!\left(t \right),  
		\hat{n}_{-k}^\nu\!\left(0\right) \right]} e^{i\omega t} dt\,,
	\end{equation}
	where $\hat{n}_{k}^\mu = \frac{1}{\sqrt{N}} \sum_p \tilde{\Psi}^\d_{p-k}\sigma^\mu\tilde{\Psi}_p$ is the Fourier transform of $\hat{n}_i^\mu$ and $*$ denotes a convolution. With $\eta$ a positive infinitesimal, the first term is the Fourier transform of $-i\Theta\left(t\right)$, which introduces the poles of $\chi$. For a time-translation invariant system the second term is self-adjoint under the transposition $\mu \leftrightarrow \nu$, implying that the susceptibility has the analytical structure $\chi\left(\omega\right)=\sum_\text{poles}\frac{\text{Hermitian matrix}}{\text{simple pole}}$. We thus define the Hermitian spectral weight function $\mathcal{A}_{\mu\nu}(k,\omega)$ by
	\begin{equation}\label{eq:chiSpectralPoleDecomposition}
	\chi_{\mu\nu}(k,\omega)=
	\int_{-\infty}^{\infty}\frac{\mathcal{A}_{\mu\nu}(k,\omega')}{\omega^+-\omega'}d\omega'
	\end{equation}
	where $\omega^+ = \omega + i\eta$. 
	The modes of the system are located where $\mathcal{A}_{\mu\nu}\left(k,\omega\right)$ is non-zero. In the excitonic insulating state we expect a continuous region of modes above the gap energy, corresponding to excitations of electrons from the valence to the conduction band. By contrast, the collective modes that lie below this particle-hole continuum will form a sharp line in the $\left(k,\omega\right)$ plane, with a dispersion relation $\omega=\omega_k$. As the collective modes are gapless, they should be separated from the continuum, at least for small momenta. 
	While Eq.~\eqref{eq:chiSpectralPoleDecomposition} is suggestive of the Kramers-Kronig relation satisfied by the retarded $\chi$ \cite{AltlandSimons2010}, we remark that it does not define $\mathcal{A}_{\mu\nu}$ uniquely. Instead, leveraging the identity $\mathrm{Im}\left(\omega^+ - \omega'\right)^{-1} = -\pi \delta \left(\omega - \omega'\right)$ along with $\mathcal{A}$'s hermiticity, we invert this definition and find
	\begin{equation} \label{eq:SpectralWeightFunctionASolution}
	\mathcal{A}_{\mu\nu}
	= 	\frac{1}{2\pi i} \left[\left( \chi_{\mu\nu} \right)^\d - \chi_{\mu\nu}  \right]\,.
	\end{equation}
	
	\heading{Collective modes}
	The expansion \eqref{eq:ChiWickTheoremConnected} begins by evaluating the Kubo formula with respect to some initial ground state ansatz $\ket{0}$ which approximates the true interacting ground state. For this we take the ground state found within the mean-field approximation. This solution is worked out in Ref.~\cite{Murakami2017} and outlined in Appendix~\ref{app:GroundStateProperties}, where we also list the resulting Feynman diagrammatic rules. 
	We employ the random-phase approximation (RPA) \cite{Bohm1953, Pines1953, Chen2017}. Within the RPA, the interacting susceptibility is given by a Dyson equation \cite{FetterWalecka}, represented schematically by
	\begin{equation}
	\underset{\chi^\text{RPA}}{\bubbleRPA} = \underset{\chi^0}{\bubble} 
	+ \underset{\,\,\chi^0 \, * \,  U^\text{eff} \, * \, \chi^\text{RPA}}{\bubble\bubbleRPA}\,.
	\end{equation}
	Here the clear bubble $\chi^0$ is the bare susceptibility, and $U^\text{eff}$ is a numerical vertex factor corresponding to the projection of the electron--electron interaction operator onto the sector or bubble diagrams.
	The convolutions are turned into simple (matrix) products in momentum space, which we will use exclusively henceforth.
	Isolating $\chi^\text{RPA}$, we obtain the familiar RPA result \cite[cf. Sec. 3]{Chen2017}
	\begin{equation} \label{eq:chiRPASolution}
	\chi^\text{RPA}_{\mu\nu}\left(k,\omega\right)
	=\chi^0_{\mu\nu}
	+\sum_{\lambda,\eta} \chi^0_{\mu\lambda} U^\text{eff}_{\lambda\eta}  \chi^\text{RPA}_{\eta\nu} 
	=\left[\frac{\chi^0}{1-U^\text{eff}\chi^0}\right]_{\mu\nu}\,.
	\end{equation}
	Here the solution for $\chi^\text{RPA}$ is shorthand notation for matrix inversion.
	Note that Eq.~\eqref{eq:chiRPASolution} is a $4\times4$ matrix equation;  $U^\mathrm{eff} = \left(U_0/2\right) \mathrm{diag}\left(1,-1,-1,-1\right)$ and $\chi^0$ are computed in Appendix \ref{app:GroundStateProperties}. For $k=0$ the components $\mu,\nu=x,\,y,\,z$ do not mix with the $\mu=\nu=0$ sector, as $\hat{n}^0_{k=0}$  is the total particle number and is conserved \cite[SI]{Murakami2017}. However, this does not hold for general $k$, and we must consider this equation in full \cite{particle-hole}.

	\begin{figure}[tb]
		\centering
		\includegraphics[width=\apsfigurewidth]{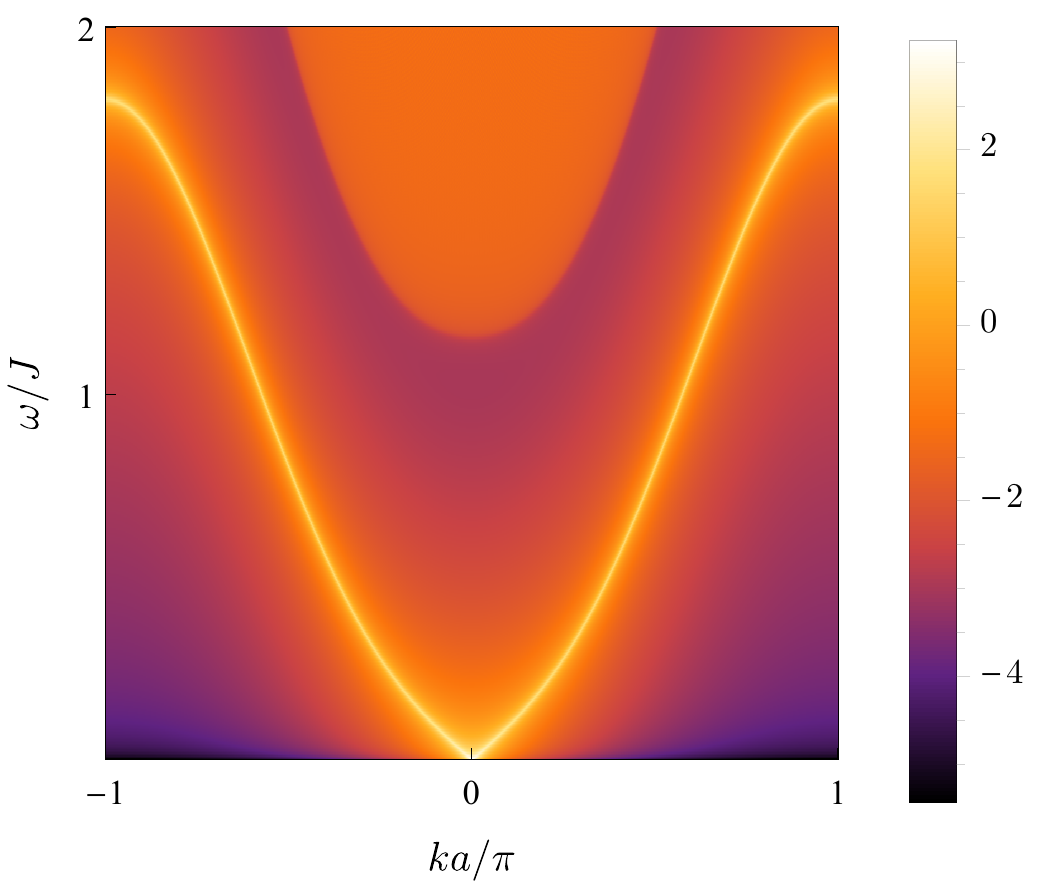}
		\caption{Component $\mathcal{A}_{yy}$ of the spectral function (\ref{eq:SpectralWeightFunctionASolution}),  corresponding to order parameter oscillations in the ``phase direction''. The colour scale is logarithmic (in units $J^{-1}$), showing the singularities at the poles; the broadening regulator used here is \revised{ $\eta = 0.005~J$.} The system size is $N = 1024$ sites. The gapped two-particle excitation continuum is visible at high energies. In the symmetry-broken EI, a Goldstone branch of CMs appears, and considerable spectral weight is shifted to it from the continuum.}
		\label{fig:Chi0andChiRPA}
	\end{figure}

	\heading{Results}
	The RPA susceptibility is evaluated and plotted in Fig.~\ref{fig:Chi0andChiRPA}. The displayed quantity is $\mathcal{A}_{yy}$, which corresponds to the spectral density of modes that oscillate in the ``phase direction'' of the order parameter. A non-zero \revised{$\eta = 0.005~J$} is used to broaden mode poles. 
	The bottom of the two-particle continuum is visible, occurring at the frequency of the inter-band gap seen in Fig.~\ref{fig:quasiparticlebands}. 
	Additionally, a separated branch of modes $\omega_k$ appears at lower energies, revealing the expected Goldstone-like branch of CMs.
	At fixed $k$, this branch absorbs a considerable portion of the spectral weight which was once in the two-particle continuum in the bare $\chi^0$. 
	The eigenvectors of $\mathcal{A}_{\mu\nu}(k,\omega_k)$ indicate the nature of the collective modes, and at small $k$ these are predominantly phase modes.

	\section{Disorder Scattering} \label{sec:Disorder}
	
	The focus of this paper is to determine the disorder-induced scattering rate of excitations. We add to the model an external disorder potential and a corresponding one-particle vertex,
	\begin{align} \label{eq:DisorderHamiltonian}
	H_\text{dis} & = \feynmandiagram [small, baseline={([yshift=-.5ex]current bounding box.center)}, layered layout, horizontal=a to b] {c -- [fermion] a -- [insertion={[style=red,size=5pt]0.5}]  b -- [fermion] d}; 
	\tcbr
	= \sum_{\lambda} \sum_{i} V^\lambda_i \Psi^\d_i\sigma^\lambda\Psi_i^{\phantom{\dagger}} 
	= \frac{1}{\sqrt{N}} \sum_{\lambda} \sum_{k,q} \tilde{V}^\lambda_q \tilde{\Psi}^\d_{k+q} \sigma^\lambda \tilde{\Psi}_k^{\phantom{\dagger}} \,,
	\end{align}
	where $V^\lambda_i$ ($\lambda = 0,\,x,\,y,\,z$) is a real random disorder field with Fourier transform $\tilde{V}^\lambda_q$. We assume that $V^\lambda$ is normally distributed with zero mean and the following covariance relations
	represented by a ``disorder propagator''
	\begin{equation} \label{eq:DisorderPotentialCovarianceMatrix}
	\feynmandiagram [small, baseline={([yshift=-0.5ex]current bounding box.center)}, horizontal=a to b] {
		a -- [red, boson] b
	};=
	\left< V^\lambda_i V^{\lambda'}_j \right>  = K_{\left|i-j\right|}^{\lambda \lambda'} \Leftrightarrow
	\left< \tilde{V}^\lambda_q \tilde{V}^{\lambda'}_{-q} \right> =  \tilde{K}^{\lambda \lambda'}_q\,.
	\end{equation}
	
	For simplicity we henceforth assume the disorder potential is delta-correlated in real space, $\braket{V^\lambda_i V^{\lambda'}_j}~\propto~\delta_{ij}$, signifying an ensemble of point scatterers. Furthermore we take the different disorder channels to be uncorrelated.  We thus set $\tilde{K}^{\lambda\lambda'}_q = K_0 \delta_{\lambda \lambda'}$. 
	
	The four disorder channels can be classified with respect to the $U(1)$ symmetry of the Hamiltonian: Channels $\lambda = 0,\,z$ preserve the symmetry, while channels $\lambda = x,\, y$ violate it. We will show that the symmetry of the disorder has crucial consequences for the scattering of the CMs. Physically, channel $\lambda = 0$ corresponds to an external electrostatic potential.
	
	Disorder will induce a finite lifetime for all excitations, which at the level of RPA were infinitely long lived.
	This will manifest in the broadening of the poles of the single- and two-particle response functions. 
	We will obtain these induced scattering rates by computing the electron and CM (proper) self-energy \cite{FetterWalecka}. 
	Working to second order in $H_\mathrm{dis}$, this is equivalent to computing lifetimes with Fermi's Golden Rule (FGR) \cite{Sakurai2017}. However, the diagrammatic method is useful for computing the scattering rate of the CMs, whose wavefunctions are not found explicitly. 
		
	\revised{Our work extends that of Zittartz \cite{Zittartz1967}, who calculated the lifetime of the EI quasiparticle excitations in the presence of disorder. The present work builds upon it in three aspects: (i) We also compute the lifetime of the collective modes. (ii) We consider all possible disorder channels;  Zittartz's ``normal impurities'' correspond to our $\lambda=0$ disorder channel. (iii) We take into account the fact that the disorder potential is dressed by the condensate (see below). This turns out to be crucial in obtaining the correct lifetimes of the collective modes when scattered by symmetry-respecting disorder, such as Zittartz's.}

	\heading{Disorder dressing}
	The inclusion of disorder modifies the ground state of the system above which the collective modes are excited. This manifests as a change in the value of the mean fields that the electronic degrees of freedom are coupled to. This modulation corresponds to the condensate shifting to screen the disorder. Assuming the disorder is perturbative, this can be accounted for to leading order by using linear response, as $\delta \braket{n^\mu}\left(q\right) = \sum_\nu \chi^\mathrm{RPA}_{\mu\nu}\! \left(q,\omega = 0\right)\tilde{V}^\nu_q$. This implies a dressed disorder vertex
	\begin{equation}
	\tilde{\mathbb{V}}_q^\lambda = \tilde{V}_q^\lambda + \sum_{\mu,\nu} U^\mathrm{eff}_{\lambda \mu} \chi^\mathrm{RPA}_{\mu\nu} \!\left(q, \omega = 0\right)\tilde{V}_q^\nu\,,
	\end{equation}
	which we absorb into a dressed disorder propagator
	\begin{align}
	& \braket{\tilde{\mathbb{V}}_q^\lambda \tilde{\mathbb{V}}_{-q}^{\lambda'}} 
	= \tilde{\mathbb{K}}_q^{\lambda \lambda'}\tcbr
	= \sum_{\mu,\nu} \left[\mathds{1} + U^\mathrm{eff}\chi^\mathrm{RPA}_{\omega=0}\right]_{\lambda \mu}\! \tilde{K}_q^{\mu\nu} \!\left[\mathds{1} + U^\mathrm{eff} \chi^\mathrm{RPA}_{\omega=0}\right]_{\nu\lambda'}^\d\,.
	\end{align}
	Diagrammatically, this is equivalent to the screening
	\begin{align}
	& \feynmandiagram [small, baseline={([yshift=-0.5ex]current bounding box.center)}, horizontal=a to b] {
		a -- [red, thick, boson] b
	};
	= \disorderprop + \disorderprop \bubbleRPA \, + \tcbr  \bubbleRPA \disorderprop + \bubbleRPA \disorderprop \bubbleRPA \,.
	\end{align}
	Though the bare disorder channels could be uncorrelated (i.e. $\tilde{K}_q$ is diagonal), in general the channels become mixed in $\tilde{\mathbb{K}}_q$. However, it can be shown [Appendix \ref{app:GroundStateProperties}] that channel $\lambda = y$ remains uncorrelated with the rest.
	The dressing of the disorder is a crucial effect if the disorder respects the {$U(1)$ symmetry of  the clean system} (shown below).

	\subsection{Electron propagator}

	\begin{figure}[tb!]
		\centering
		\includegraphics[width=\apsfigurewidth]{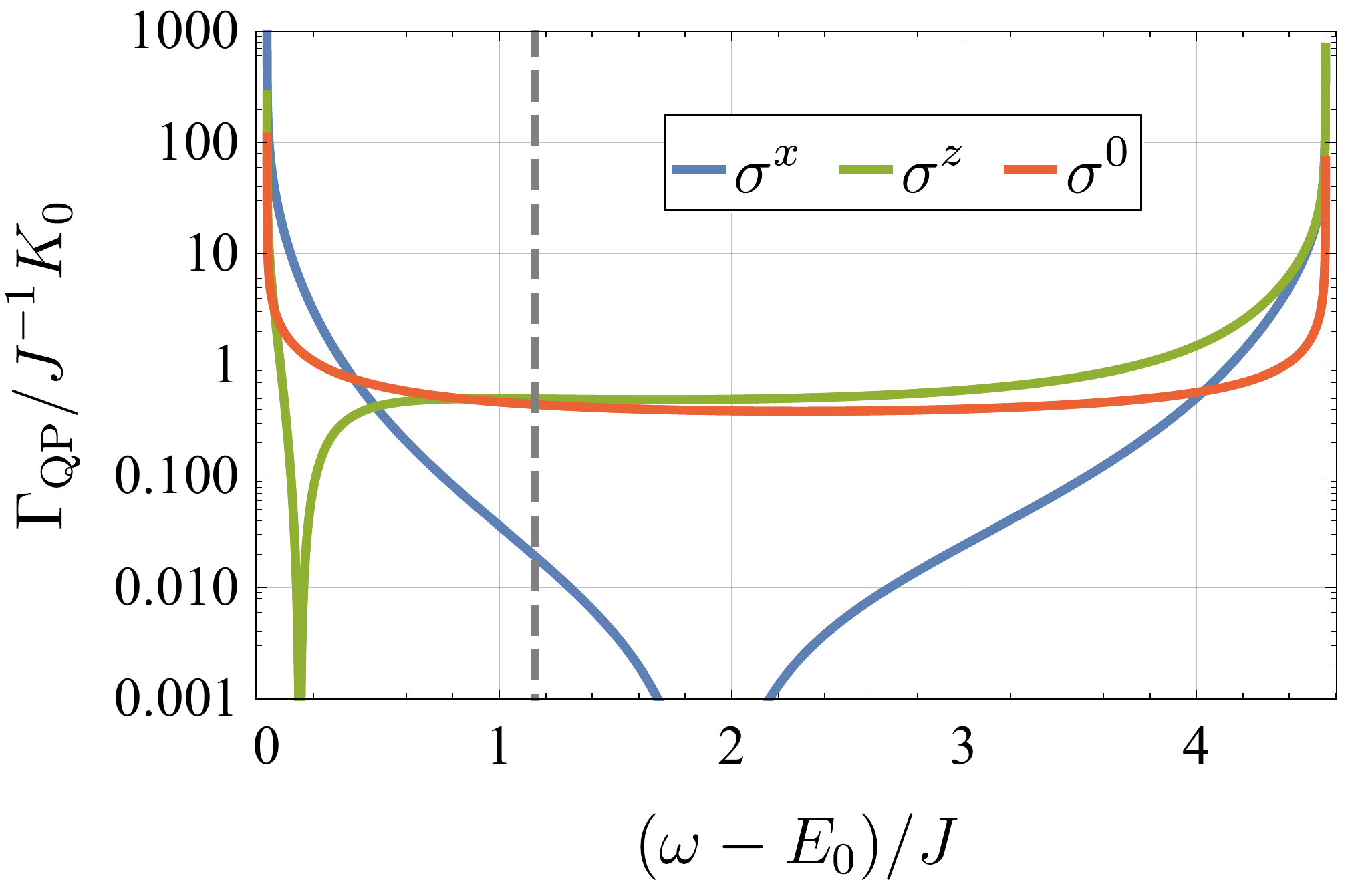}
		\caption{Single-particle scattering rates, for different scattering channels. 
			The mode energy is measured relative to the bottom of the conduction band \revised{$E_{k=0}$}.
			The peaks at the band edges are van Hove singularities in the one-dimensional density of states. 
			The features in channels $\lambda=x,z$ are explained in the text. The vertical dashed line marks the bottom of the electron--hole two-particle continuum, for reference.
			(Here $N = 1024$.)}
		\label{fig:SingleParticleDecayRates}
	\end{figure}

	\heading{Quasiparticle scattering rates}
	The dressed propagator can be determined by the usual Dyson equation   
	\begin{equation} \label{eq:SingleParticleDysonEquation}
	G = G^0 + G^0 \Sigma G 
	\,\Leftrightarrow\,
	G^{-1} = \left(G^0\right)^{-1} - \Sigma\,,
	\end{equation}
	with $\Sigma$ the proper self-energy insertion \cite{FetterWalecka}. Working to second order in $V^\lambda$, it is found by evaluating the diagram 
	\begin{equation} \label{eq:SingleParticleSelfEnergyDiagram}
	\Sigma\left(k,\omega\right) = -i \times
	\feynmandiagram [inline=(b.base),layered layout, horizontal=a to b] {
		a -- [fermion] b -- [boson, red, thick, half right] a
	};
	= 
	\frac{1}{N} \!\! \sum_{\lambda, \lambda',q} \!\! \tilde{\mathbb{K}}_q^{\lambda\lambda'} \sigma^{\lambda} G^0 \!\left(k + q,\omega\right) \sigma^{\lambda'}.
	\end{equation}
	As the disorder propagator is statistical rather than quantum-mechanical, tadpole diagrams are considered disconnected and cancel by the linked cluster theorem.
	
	Furthermore, $G^0\left(k,\omega\right)$ is diagonalized by a unitary matrix $R_k$ which rotates it to the basis of the hybridized quasiparticle excitations $\phi^\d_\pm\left(k\right)$ [Appendix~\ref{app:GroundStateProperties}], where it takes the form
	\begin{equation} \label{eq:SingleParticleBareGDiagonalized}
	G^0\left(k,\omega\right) = R_k \begin{pmatrix}
	\omega^+ - E_+(k) & \\
	& \omega^-  - E_-(k)
	\end{pmatrix}^{-1} R_k^\d \,.
	\end{equation}
	Thus, Eq.~\eqref{eq:SingleParticleDysonEquation} reads
	\begin{equation}
	G^{-1} = R_k \begin{pmatrix}
	\omega^+ - E_+  - \tilde{\Sigma}_{11} &  -\tilde{\Sigma}_{12} \\
	-\tilde{\Sigma}_{21} & \omega^-  - E_- - \tilde{\Sigma}_{22}
	\end{pmatrix} R_k^\d \,
	\end{equation}
	where $\tilde{\Sigma}\left(k, \omega\right) = R_k^\d \Sigma\left(k,\omega\right) R_k$ is  the self-energy ``rotated'' to the same eigenbasis. The poles of the dressed propagator are where it is singular, i.e. $\omega=E_k$ at which $\det{G^{-1}\left(k, E_k\right)}=0$. The off-diagonal elements of $\tilde{\Sigma}$ only come into play in post-leading contributions to the new $E_k$. Focusing henceforth on the positive energy poles, it is now shifted and broadened by 
	\begin{equation} \label{eq:SingleParticleSelfEnergyRate}
	\delta E_k = \mathrm{Re}\,{\tilde{\Sigma}_{11}\!\left(k,E_k\right)}, \quad 
	\Gamma^\text{QP}_k = - \mathrm{Im}\,{\tilde{\Sigma}_{11}\!\left(k,E_k\right)}.
	\end{equation} 
	
	The matrices $R_k$ are real, 
	\begin{equation}
	R_k = \exp\left({\frac{i}{2}\sigma^y\theta_k}\right) = 
	\begin{pmatrix}
	\phantom{-}	\cos{\frac{\theta_k}{2}}	\,& \sin{\frac{\theta_k}{2}} \\ 
	-		\sin{\frac{\theta_k}{2}}	\,& \cos{\frac{\theta_k}{2}}
	\end{pmatrix}
	\end{equation}
	where the mixing angle $\theta_k$ which hybridizes the original bands is the polar angle of the pseudomagnetic field $\vec{B}_k=\left(-\sin{\theta_k},0,\cos{\theta_k}\right)|B_k|$ obtained in the mean-field solution [Appendix~\ref{app:GroundStateProperties}]. They thus do not mix the real and imaginary parts of the self-energy.

	The self-energy diagram is computed in Appendix \ref{app:QuasiparticleSelfEnergy}. We find the scattering rate to be 
	\begin{equation} \label{eq:SingleParticleFermiGoldenRuleRateBothRates}
	\Gamma^{\mathrm{QP}}_k \!\!
	=\! \frac{\pi}{2}  g\!\left(E_k\right)  \revised{  \langle \Big| \! \sum_\lambda \! \tilde{\mathbb{V}}_{-2k}^{\lambda} \hat{B}_{k}^{\lambda} \Big|^2 \rangle  \!}
	= \! \frac{\pi}{2}  g\!\left(E_k\right) \! \sum_{\lambda,\lambda'} \! \hat{B}^\lambda_k  \tilde{\mathbb{K}}_{-2k}^{\lambda \lambda'} \hat{B}^{\revised{\lambda'}}_k,
	\end{equation}
	where $\hat{B}_k^0 = 1$. Here we have put the scattering rate into a form recognizable as FGR \cite{Sakurai2017}, and $g\left(E\right)$ is the single-particle density of states (DoS) per site. 
	The last inner product is the matrix element for elastic backward scattering with momentum transfer $q = -2k$. {Here we discarded the $q=0$ matrix element, as forward scattering does not affect transport.}
	Note that $\hat{B}_k^y=0$. As the $\tilde{\mathbb{K}}_q^{yy}$ element is not mixed with other elements in the screening process, we find that the $\lambda=y$ channel has a vanishing scattering rate.
	
	\revised{This result qualitatively reduces to that of Zittartz (cf. Eq.~(31) in Ref.~\cite{Zittartz1967}) if we only consider the electrostatic channel ($\lambda=\lambda'=0$), and neglect the disorder dressing (i.e. use $K^{00}$ instead of $\mathbb{K}^{00}$). However, the dressed $\lambda=0$ channel generally mixes with additional channels, giving rise to a momentum dependence other than through the density of states.}
	
	\heading{Results}
	Eq.~\eqref{eq:SingleParticleFermiGoldenRuleRateBothRates} is evaluated and the contribution of each channel $\lambda$ is plotted in Fig.~\ref{fig:SingleParticleDecayRates}. \revised{The quasiparticle energies are measured relative to the bottom of the (interacting) conduction band $E_{k=0}$.} The sharp peaks at the band edges are the one-dimensional van Hove singularities in the DoS.  
	It is apparent that the different scattering channels induce scattering rates of the same order of magnitude at most energies. The rate \eqref{eq:SingleParticleFermiGoldenRuleRateBothRates} can be interpreted as the inner product between the disorder potential and the pseudomagnetic field. The screening ``rotates'' the disorder channels, such that the potential may be orthogonal to the field at specific $k$ points, and this is the origin of the dips in channels $\lambda = x, z$. It is curious that this allows the backscattering rate to vanish. However, this is an artifact of one spatial dimension, as in higher dimensions angular summation over different momentum transfers will produce a non-zero rate.
	\subsection{Density-density response function}
	
	\heading{Diagrammatic expansion}
	The CM scattering rates may be found by resumming diagrammatic insertions to obtain the leading-order contributions to the exciton self-energy. For this to yield the scattering rates, we need to find an effective self-energy which satisfies a Dyson-type equation, that is schematically
	\begin{equation} \label{eq:ChiDysonExpansionSimple}
	\chi^\text{dis} = \chi^\text{RPA} + \chi^\text{RPA} * \Sigma_\text{eff} * \chi^\text{dis} 
	= \left[\left(\chi^\text{RPA}\right)^{-1} - \Sigma_\text{eff}\right]^{-1} \,.
	\end{equation}
	We enumerate all the RPA-type second-order disorder diagrams which enter expansion \eqref{eq:ChiWickTheoremConnected}, which are 
	\begin{widetext}
		\begin{equation} \label{eq:DisorderDiagrams}
		\Sigma \,=\, \bubbleDisordered \,=\, \bubbleClam \,+\, \bubblePurse \,+\, 
		\begin{tikzpicture}[baseline={([yshift=-.5ex]m.base)}]
		\begin{feynman}
		\vertex (m) at (0,0) ;
		
		\vertex (a) at (-\bubblesize, 0);
		\vertex (b) at (+\bubblesize,0);
		
		\vertex (u1) at (-0.5*\bubblesize,\bubblesize*0.414213*0.5);
		\vertex (u2) at (+0.5*\bubblesize,\bubblesize*0.414213*0.5);
		
		\diagram* {
			{%[edge=fermion]
				(a) -- [out = 45, in = 180] (u1) -- [out = 0, in = 135] (m) -- [out = -135, in = -45] (a),
				(m) -- [out = 45, in = 180] (u2) -- [out = 0, in = 135] (b) -- [out = -135, in = -45] (m)},
			(u1) -- [boson, red, thick, quarter left] (u2),
		};
		\end{feynman}
		\end{tikzpicture}
		\,+\,
		\begin{tikzpicture}[baseline={([yshift=-.5ex]m.base)}]
		\begin{feynman}
		\vertex (m) at (0,0) ;
		\vertex (n) at (+\bubblesize,0) ;
		
		\vertex (a) at (-\bubblesize, 0);
		\vertex (b) at (2*\bubblesize,0);
		
		\vertex (u1) at (-0.5*\bubblesize,\bubblesize*0.414213*0.5);
		\vertex (u2) at (+1.5*\bubblesize,\bubblesize*0.414213*0.5);
		
		\diagram* {
			(m) -- [fermion, quarter left]  (n) -- [fermion, quarter left] (m),
			{%[edge=fermion]
				(a) -- [out = 45, in = 180] (u1) -- [out = 0, in = 135] (m) -- [out = -135, in = -45] (a),
				(n) -- [out = 45, in = 180] (u2) -- [out = 0, in = 135] (b) -- [out = -135, in = -45] (n)},
			(u1) -- [boson, red, thick, quarter left] (u2),
		};
		\end{feynman}
		\draw [pattern=north east lines] (m) arc [start angle=135, end angle=45, radius=\bubblesize/sqrt(2)] arc [start angle=-45, end angle=-135, radius=\bubblesize/sqrt(2)];
		
		\end{tikzpicture}\,.
		\end{equation}
		%\end{widetext}
		Here the direction of the fermion propagators is left unspecified to indicate that the disorder potential could couple to either the electron or hole propagator in each bubble; therefore, the diagrams have 1, 2, 4, and 4 realizations, respectively. The different diagrams in Eq.~\eqref{eq:DisorderDiagrams} are computed and their properties discussed in Appendix \ref{app:ExcitonSelfEnergyDiagrammatics}. 
		
		We can now systematically generate the diagrammatic expansion for the disordered RPA susceptibility $\chi^\text{dis}$ by concatenating arbitrary numbers of clean and disordered bubble diagrams. This is represented schematically by
		%\begin{widetext}
		\begin{align} \label{eq:DisoerderResummation}
		i \chi^\text{dis} & \,=\, \bubbleRPADisordered \,=\, 
		\bubbleRPA \,+\, \bubbleDisordered \nonumber % \\ & 
		\,+\, \bubbleRPA\bubbleDisordered \,+\, \bubbleDisordered\bubbleRPA \,+\, %\bubbleDisordered\bubbleDisordered \,+\, 
		\dots \nonumber \\ &
		= \bubbleRPA + \Big(1+\bubbleRPA\Big)\bubbleDisordered
		\sum_{n=0}^\infty \Bigg[\Big(1+\bubbleRPA\Big)\bubbleDisordered\Bigg]^n
		\Big(1+\bubbleRPA\Big) \nonumber \\ &
		= \Bigg[\bubble \,+\, \bubbleDisordered \Bigg] \Bigg[1 \,+\, \bubbleRPADisordered\Bigg]\,.
		\end{align}
	\end{widetext}
	Note that two consecutive clean RPA bubbles are not allowed, as this leads to double counting. Therefore, the Dyson relation in the last line displays the bare $\chi^0$ instead of $\chi^\text{RPA}$, and it cannot be used to extract the effective self-energy in Eq.~\eqref{eq:ChiDysonExpansionSimple}.
	
	Instead, we focus on the second line of Eq.~\eqref{eq:DisoerderResummation}, which reads
	\begin{align} \label{eq:DisorderSelfEnergyDefinition}
	\chi^\text{dis} & = \chi^\text{RPA} + \left(1 + \chi^\text{RPA} * U^\text{eff}\right) * \Sigma ~ * \tcbr 
	\sum_{n=0}^\infty \left[ \left(  1 + U^\text{eff} \!*\!  \chi^\text{RPA} \right) 
	\! * \! U^\text{eff} \!*\! \Sigma  \right]^n \!\! * 
	\left(1 + U^\text{eff} \!*\! \chi^\text{RPA} \right).
	\end{align}
	Adding $\left(U^\text{eff}\right)^{-1}$ to both sides and regrouping products, we can rewrite
	\begin{equation} \label{eq:EffectivePropagator}
	\mathcal{G}^\text{dis} = 
	\mathcal{G}^\text{RPA} + 
	\mathcal{G}^\text{RPA} * \Sigma^\star * \mathcal{G}^\text{RPA}
	\end{equation}
	with $\mathcal{G}^\text{dis, RPA} = \left(U^\text{eff}\right)^{-1} + \chi^\text{dis, RPA}$ and the self-energy insertion
	\begin{equation} \label{eq:ExcitonImproperSelfEnergy}
	\Sigma^\star = \left[U^\text{eff} * \Sigma * U^\text{eff} \right] * 
	\sum_{n=0}^{\infty} \left[ \mathcal{G}^\text{RPA}*
	\left[  U^\text{eff} * \Sigma * U^\text{eff}  \right]\right]^n\,.
	\end{equation}
	
	This solution illustrates the analogy between $\chi$ and the exciton propagator: 
	From Eqs.~\eqref{eq:EffectivePropagator} and \eqref{eq:ExcitonImproperSelfEnergy} we identify $\mathcal{G}^\text{RPA,dis}$ as the ``bare'' and ``dressed'' effective exciton propagators, respectively, and
	$\Sigma^\star$ as the \emph{improper} self-energy insertion \cite[Sec. 9]{FetterWalecka}. 
	The effective \emph{proper} self-energy is then read from the definition 
	$\Sigma^\star = \Sigma_\text{eff} * \sum_n \left[ \mathcal{G}^\text{RPA} * \Sigma_\text{eff} \right]^n$. 
	We thus conclude
	\begin{align}
	\mathcal{G}^\text{dis} & 
	= \mathcal{G}^\text{RPA} \!+ \mathcal{G}^\text{RPA} \!* \Sigma_\text{eff} * \mathcal{G}^\text{dis} 
	=\left[\left(\mathcal{G}^\text{RPA}\right)^{-1} \!- \Sigma_\text{eff}\right]^{-1}\!\!,
	\\
	\Sigma_\text{eff} & = U_\text{eff} * \Sigma * U_\text{eff}\,. \label{eq:ExcitonEffectiveSelfEnergy}
	\end{align}
	
	Note that $U^\text{eff}$ is regular, and therefore $\mathcal{G}$ has no other poles except those of $\chi$. Additionally, $U^\text{eff}$ is itself Hermitian, and therefore we can substitute $\mathcal{G}$ into Eq.~\eqref{eq:SpectralWeightFunctionASolution} instead of $\chi$ without affecting $\mathcal{A}$.
	
	\heading{Scattering rates}
	Given the effective excitonic self-energy, we now need to extract from it the collective mode scattering rates in a procedure analogous to the ``rotation'' used to extract the single-particle rates. 
	
	We recall that the collective modes are located where $\mathcal{A_{\mu\nu}}(k,\omega)$ is non-zero. 
	Specifically, each collective mode at $\left( k, \omega_k \right)$ is associated with an eigenvector of $\mathcal{A_{\mu\nu}}\left( k, \omega_k \right)$. Although it is a $4 \times 4$ matrix, only eigenvectors with non-zero eigenvalues represent physical modes. 
	Thus, we decompose the weight function as
	\begin{align} \label{eq:SpectralADecomposition}
	\mathcal{A} \!\left( k, \omega \right) & = 
	\delta \! \left( \omega - \omega_k \right) \mathcal{R}_k 
	\begin{pmatrix}
	\hspace{0.00cm} \mathcal{A}_1(k) \hspace{1.50cm} \\ 
	\hspace{0.50cm} \mathcal{A}_2(k) \hspace{1.00cm} \\ 
	\hspace{1.00cm} \mathcal{A}_3(k) \hspace{0.50cm} \\ 
	\hspace{1.50cm} \mathcal{A}_4(k) \hspace{0.00cm}
	\end{pmatrix} \mathcal{R}_k^\d \tcbr + \text{sum over continuum modes}\,,
	\end{align}
	where the diagonalizing matrices $\mathcal{R}_k$ are defined such that the real eigenvalues $\mathcal{A}_i\!\left(k\right)$ are ordered by decreasing magnitude. Furthermore, we only expect one soft mode, and thus that only $\mathcal{A}_1$ is non-zero. {In Appendix \ref{app:ExcitonSelfEnergyDiagrammatics} we show that $\mathcal{A}_1 \sim \omega_k^{-1}$ for small $k$.}
	Since the continuum modes are separated from the collective modes, we will ignore their contribution. Equally we ignore the sharp CM poles at negative energies and focus on $\omega > 0$.
	
	Plugging this into Eq.~\eqref{eq:chiSpectralPoleDecomposition}, close to a collective mode pole $\left(k, \omega_k\right)$ of the clean system we have	a pole of the form
	\begin{equation}
	\left[\mathcal{G}^\text{dis}( k, \omega )\right]^{-1}  \sim 
	\mathcal{A}_1^{-1} \!\left(k\right) \left[ \omega^+ - \omega_k - \mathcal{A}_1\!\left(k\right) \tilde{\Sigma}_\text{eff} \left( k, \omega_k \right) \right],
	\end{equation}
	again with a ``rotated'' self-energy $\tilde{\Sigma}_\text{eff} \left(k, \omega \right)= \big[\mathcal{R}^\d_k \Sigma_\text{eff} \left( k, \omega \right) \mathcal{R}^{\phantom{\d}}_k\big]_{11}$. The scattering rate is then identified as 
	\begin{equation} \label{eq:ExcitonDecayRate}
	\Gamma^\text{CM}_k = -\mathcal{A}_1 \! \left( k \right)   \mathrm{Im}\left[ \tilde{\Sigma}_\text{eff} \! \left( k, \omega_k \right)\right]\,.
	\end{equation}
	\revised{Here we evaluate the self-energy at the clean pole $\omega_k$, as the shift in the pole frequency due to $\mathrm{Re}[\tilde{\Sigma}_\mathrm{eff}]$ contributes a sub-leading correction. In this we assumed that the disorder is weak such that this shift is much smaller than $\omega_{k}$. Therefore, this treatment applies to modes at frequencies $\omega_{k}$ that are larger than the gap opened by introducing a symmetry-violating disorder potential.}
	
	\begin{figure}[tb]
		\centering
		\includegraphics[width=\apsfigurewidth]{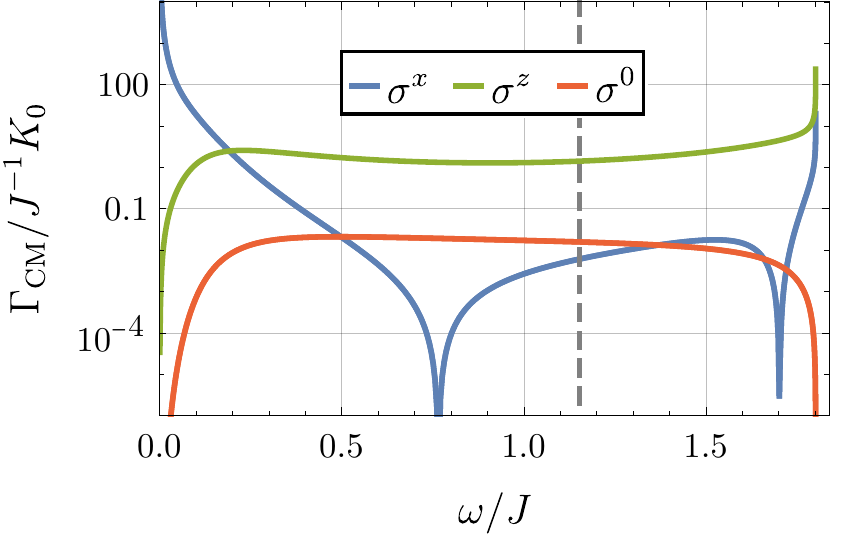}		
		\caption{Disorder-induced scattering rates of the collective modes, for different scattering channels. The scattering rate for channel $\sigma^y$ is zero at the resolution of our calculation. The dashed line marks the bottom of the two-particle continuum. Above the two-particle continuum gap, the calculation is impacted (and for channel $\sigma^0$, is dominated) by additional diagrams, which represent an exciton scattering into the two-particle continuum, i.e., disassociating; this contribution is omitted. The dips in channel $\lambda = x$ are analogous to the vanishing backwards scattering rate of the quasiparticles, cf. Fig.~\ref{fig:SingleParticleDecayRates} and surrounding discussion.}
		\label{fig:CollectiveModeDecayRates}
	\end{figure} 
	
	\heading{Low-energy regime}
	We will mostly be interested in the low-energy, long-wavelength dynamics of the CMs. If we restrict our treatment to energies below the particle--hole continuum, such that a CM cannot disassociate into an unbound electron--hole pair, the CM scattering rates can be considerably simplified. In Appendix \ref{app:ExcitonSelfEnergyDiagrammatics}, we show that in this regime only the last diagram in Eq.~\eqref{eq:DisorderDiagrams} affects the mode lifetime. Furthermore, this diagram can be evaluated explicitly at energies below the gap, and in analogy to Eq.~\eqref{eq:SingleParticleFermiGoldenRuleRateBothRates} we obtain a FGR-like expression for the CM scattering rate in terms of an effective CM vertex function [Eq.~\eqref{eq:CollectiveModeFermiGoldenRule1Dimension}]. This allows us to isolate the backwards scattering rate and discard the forward scattering rate, similar to our treatment of the quasiparticles. We refer to Appendix \ref{app:ExcitonSelfEnergyDiagrammatics} for the full calculation.
	
	\heading{Results} 
	We evaluate the scattering rates of the CMs numerically and plot them in Fig.~\ref{fig:CollectiveModeDecayRates}. The scattering rate in channel $\lambda = y$ is zero at the resolution of our calculation, and therefore is not shown --- this is elucidated in the next section. The dashed vertical line marks the bottom of the two-particle continuum, above which additional scattering processes become available (not shown here). Note the qualitative difference between channel $\lambda = x$, in which the scattering rate diverges at $\omega = 0$, despite a finite density of states, and channels $\lambda = 0 , z$, in which the rate vanishes at $\omega = 0$

	\section{Discussion} \label{sec:Discussion}

	\begin{figure}[bt]
		\centering
		\includegraphics[width=\apsfigurewidth]{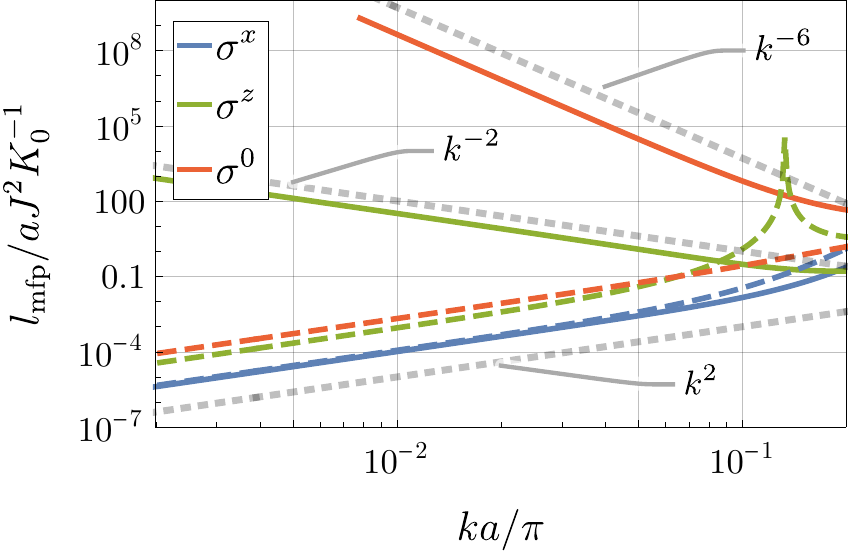} 
		\caption{Comparison between the mean free paths of the single-particle (dashed) and collective (solid) excitations, for small wavenumbers. The dotted grey lines mark the expected power-law scaling of the rate in each channel (see body of the text).		}
		\label{fig:FreePathComparison}
	\end{figure}

	\heading{Mean free paths at $k\rightarrow0$}
	To compare the transport properties of the collective modes and the quasiparticles, we consider each excitation's mean free path, defined by $l_\mathrm{mfp} = v_g/\Gamma$, with $v_g$ the group velocity and $\Gamma$ the disorder-induced scattering rate.	We focus on the dynamics at long wavelengths where clear qualitative difference arise. 
	
	Fig.~\ref{fig:FreePathComparison} shows the mean free paths of the quasiparticle excitations (dashed curves) and collective modes (solid curves) for small $k$. 
	For the quasiparticles, one sees that there are similar scattering rates for all disorder channels. This may be understood as follows: The single-particle free paths tend to zero when $k\rightarrow 0$, because of both a vanishing group velocity and a diverging DoS. If the scattering matrix element at $k = 0$ is non-zero, then at small momenta it can be treated as constant, and as in one dimension the DoS is inversely proportional to the group velocity, we find $l_\mathrm{mfp} \propto v_g^2 \sim k^2$.
	
	In contrast, with the collective modes we can identify three distinct behaviours at small $k$: (i) $l_\mathrm{mfp} \sim k^2$ in channel $\lambda = x$, similar to the quasiparticles; (ii) $l_\mathrm{mfp} \sim k^{-2}$ in channel $\lambda = z$; and (iii) $l_\mathrm{mfp} \sim k^{-6}$ in channel $\lambda = 0$. We stress that the latter two are obtained only if we include the effect of disorder screening, as discussed in Sec.~\ref{sec:Disorder}, showing that this is a crucial effect. As we now explain, this trimodality is related to the symmetry of each disorder channel: channel $x$ does not respect any of the symmetries of model \eqref{eq:Hubbard_Hamiltonian}, channel $z$ preserves its $U(1)$ condensate symmetry, and channel $\lambda=0$ also has manifest particle--hole symmetry. 
	
	\revised{We remark that these scaling laws are cut off at very small $k$, when the mode energy $\omega_k$ becomes comparable to a possible CM gap, whether intrinsic (in the case of an approximate symmetry) or induced by disorder.}
	
	\heading{Effective field theory}
	The properties that we have computed at long wavelengths can be interpreted within an effective field theory. This also allows the results to be extended to arbitrary dimension $D$.	For this purpose we consider a classical field theory for the order parameter of the condensate. To accommodate the three disorder channels discussed so far we must consider a vector order parameter; the natural field to examine is the pseudo-spin $\vec{n}_i = \frac{1}{2}\Psi^\d_i \vec{\sigma} \Psi^{\phantom{\d}}_i$ at each site.
	Taking a continuum limit, we write down the Hamiltonian density
	\begin{align}
	\mathcal{H} = & \mu n_z + \frac{1}{2} \Lambda n^2_z + \frac{1}{2} \alpha_{xy} \left[\left(\del n_x \right)^2 + \left(\del n_y \right)^2\right] \tcbr + \frac{1}{2} \alpha_z \left(\del n_z \right)^2  + \vec{V}\!\left(x\right)\cdot \vec{n}\,.
	\end{align}
	The symmetry of model \eqref{eq:Hubbard_Hamiltonian} is represented by $U(1)$ rotational symmetry around $n_z$.
	
	We may solve for the collective modes and compute the scattering rates of these modes in the face of a disorder potential; this is worked out in Appendix \ref{app:CollectiveModeFieldTheory}. There we show that the low-$k$ scaling of the rates in channels $x$ and $z$ are $k^{D-3}$ and $k^{D+1}$, agreeing with the results obtained above numerically for $D = 1$. Crucially, the qualitative difference between the channels is that at small $k$ the CMs are predominantly phase modes, which do not couple directly to the disorder channel $z$, as it respects the $U(1)$ symmetry, while it may couple to channel $x$ directly. Instead, the only mechanism which couples the phase to channel $z$ is the modulation of the ground state amplitude field configuration, which shifts to screen the disorder. 
	This mechanism also applies to channel $x$, but it is dominated by the direct coupling mechanism. Therefore, $V^z$  manifests as a spatial modulation of the coefficients in the CM wave equation without generating additional terms. Scattering in channel $\lambda = z$ is thus effectively the CM equivalent of Rayleigh scattering and scales as $\lambda^4$ in three dimensions
	
	The above field theory also predicts that channel $y$ does not induce any scattering, which is a result of arbitrarily fixing the symmetry-broken order parameter to be real. Physically, it reflects the surprising fact that unlike modulations in the stationary order parameter amplitude, modulations in its stationary phase do not scatter the collective modes. 
	
	Finally, let us address channel $\lambda=0$. While the phenomenological model above does not include this channel explicitly, we note that the same mechanism applies: As channel $\lambda = 0$ respects the condensate symmetry, it would not introduce any symmetry-violating terms in the field theory action. It may therefore do one of three things: (i) Introduce higher-order, symmetry-respecting terms, which we are not interested in, (ii) effectively contribute to $V^z$ (but crucially, not to $V^x$ or $V^y$), or (iii) spatially modulate the model parameters.  {As discussed above, the latter two options are equivalent, and thus yield the same scaling.} 	Therefore, the last ingredient required is to know how $V^0$ couples into the action. As $V^0$ corresponds to an electrostatic potential, we know on physical grounds that a constant potential would not affect the system, so $V^0$ may not appear directly and the lowest-order scalar to consider is $\del^2 V^0$. This gives us the additional $(k^2)^2$ factor between the scaling relation of channels $\lambda = z$ and $\lambda=0$. {We  demonstrate this explicitly within our lattice model in Appendix \ref{app:ExcitonSelfEnergyDiagrammatics}.}

	{This field-theoretical description allows us to readily study  the case when the $U(1)$ symmetry is only approximate. The CM at $k = 0$ then gains a gap of a magnitude $\delta$ that depends on the degree of symmetry breaking. The long-wavelength dynamics are therefore affected, the CM group velocity now vanishing at $k=0$. For cases in which the excitonic description of the insulator is a good starting point, this symmetry violation is weak such that the gap is smaller that the two-particle continuum gap, $\delta \ll \Delta$. Since disorder only induces elastic scattering, the properties of the collective modes that we have derived should all apply for CMs in the energy window  $\delta \lesssim \omega \lesssim \Delta$.}

	\heading{Experimental consequences} 
	We now turn to discuss some of the possible experimental  consequences of the low-energy collective modes of the excitonic insulator. 
	Insights from the field theory above allow us to deduce both thermodynamic and transport properties of the system at low temperatures.
	
	Firstly, the collective branch below the transition supplies additional degrees of freedom, which should add to the system specific heat at $T$ below the gap. The specific heat can be computed similarly to that of acoustic phonons, and at low temperatures we expect the usual result of Debye {\cite{Kittel}},  $C_V \sim \left(T/\Theta_\mathrm{CM}\right)^D$ where $D$ is the dimension of the system. Here $\Theta_\mathrm{CM}$ is an ``excitonic Debye temperature'' of the order of the CM band width. 
	{The relative size of this contribution compared to that of acoustic phonons with (phonon) Debye temperature $\Theta_\mathrm{D}$ is therefore of the order of $\left(\Theta_\mathrm{D}/\Theta_\mathrm{CM}\right)^D$. It is therefore likely to be most apparent in situations in which  the excitonic ordering temperature, and therefore $\Theta_\mathrm{CM}$ are small. However, the contribution can be comparable to the observed anomalies in specific heat seen in a candidate high-transition-temperature excitonic insulator \cite{Lu2017}.}

	Secondly, the collective modes have potentially very striking consequences for the transport properties of the EI. We have shown that at large wavelengths, it is possible for CMs to have large mean-free paths, if disorder in the system is symmetry-preserving. 
	These large mean free paths, combined with the electronic-scale group velocity, imply that they could have a dominant signature in the material thermal conductivity $\kappa \propto C_V v_g l_\mathrm{mfp}$. At temperatures below the particle-hole gap,
	one expects the excitonic insulator not just to transport heat via phonons, but to have a large contribution from the collective modes (at least at temperatures above any symmetry-breaking gap).
	
	If the only scattering mechanism present is disorder, the qualitative differences in the scattering of the CMs by the various disorder channels manifest as a stark differences in the resulting thermal conductivity, which we compute in Appendix \ref{app:CollectiveModeThermodynamics}. For disorder that couples to the excitonic order parameter ($\lambda = x$), $l_\mathrm{mfp} \rightarrow 0 $ for vanishing $k$, and we obtain $\kappa \sim T^3$ in all dimensions. If only the symmetry-respecting channels are present ($\lambda =0,z$), then the scattering rate vanishes for $k \rightarrow 0$. Similar to the treatment of acoustic phonons \cite{Kittel}, we can truncate the mean free path with the system linear size $L$. Therefore, the conductivity follows the specific heat and $\kappa \sim L T^D $ at very low temperatures, before plateauing once all the system-size-limited CMs are saturated. 
	
\begin{figure}[tb]
		\centering
		\includegraphics[width=\apsfigurewidth]{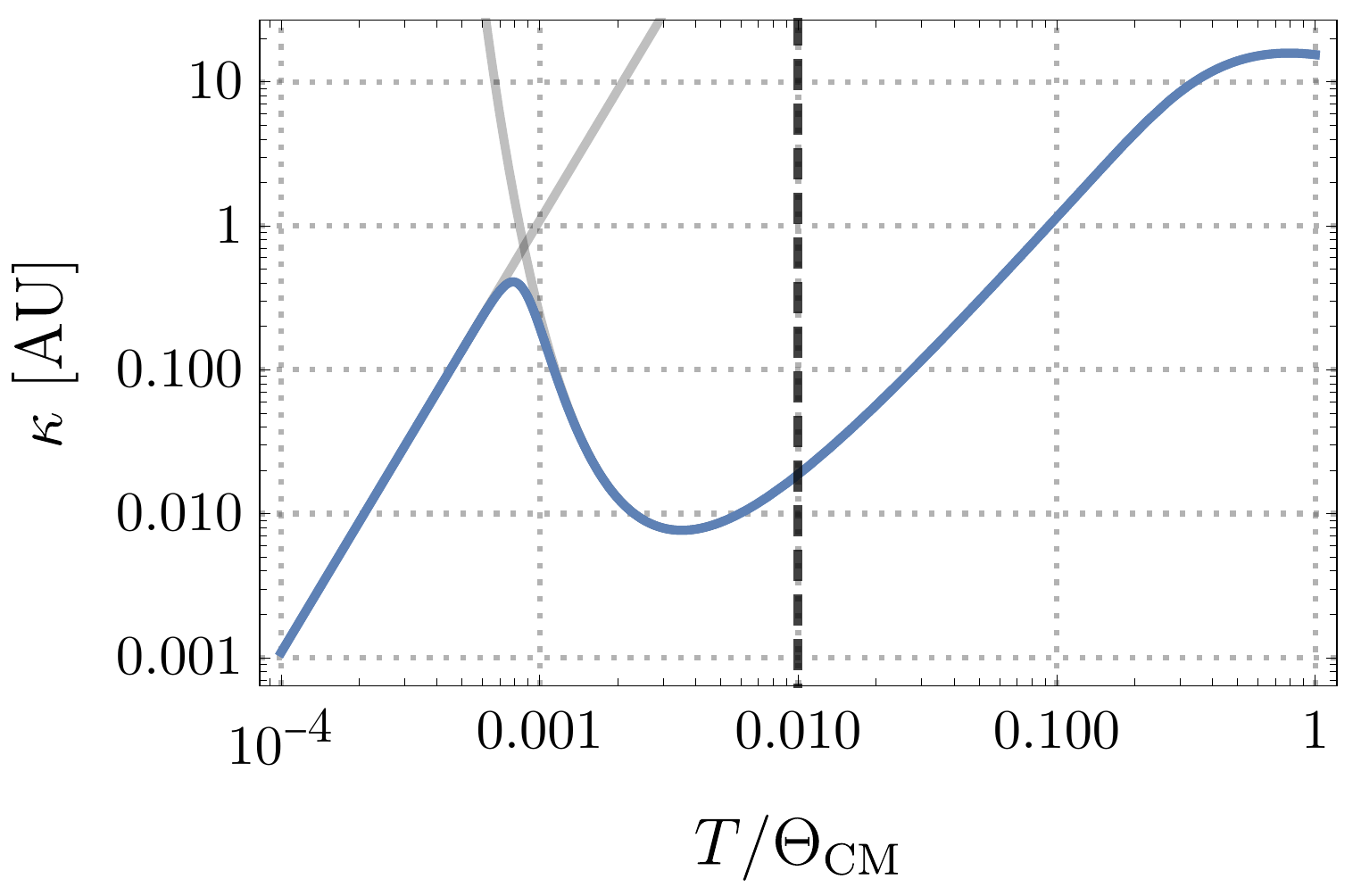}
		\caption{CM thermal conductivity as a function of temperature, when coupled to optical phonons. The vertical dashed line  marks the phonon Einstein temperature. Below it, phonon thermal excitation is suppressed, and the CM mean free path becomes dominated by the system size.}
		\label{fig:thermal_conductivity}
	\end{figure}

	In practice, as the temperature is increased, CMs will collide with thermally excited particles such as electrons, phonons, and each other. The separation of energy scales between CMs and phonons could  give rise to an interesting signature. To demonstrate this, we consider the CMs to be coupled to optical phonons with Einstein temperature $\Theta_\mathrm{E}$. The phonons are slower degrees of freedom, and on electronic timescales could be considered as static disorder fields. If the CM--phonon coupling mechanism breaks the $U(1)$ symmetry, then this scattering process would dominate all others once a thermal phonon population is excited at temperatures above $\Theta_\mathrm{E}$. However, below this temperature, mean free paths are limited by the system size and the thermal conductivity would increase. A schematic plot in such a system is depicted in Fig.~\ref{fig:thermal_conductivity}. 
	
	{Measurements of thermal conductivities of EI candidates are scarce. However, in \ce{TmSe_{0.45}Te_{0.55}} the suspected EI transition can be driven by varying pressure \cite{Wachter2004}, admitting comparison between the normal and EI phases at fixed temperatures. Indeed, below $20~\mathrm{K}$ the EI phase exhibits thermal conductivity orders of magnitude larger than that in the normal phase. Additionally, in the EI phase the thermal conductivity increases with decreasing temperature.}	This anomalous behaviour is attributed to superfluidity in the condensate \cite{Wachter2004, Wachter2005}. 
	However, it might be explained by a scenario such as that in Fig.~\ref{fig:thermal_conductivity}, in which decreasing $T$ reduces the scattering rate, thereby increasing thermal conductivity without invoking superfluidity. 	
	
	Finally, we point out that the robust collective modes could have other striking properties in dynamical measurements. The EI transition in candidate materials is often accompanied by other transitions such as a charge density wave \cite{Kogar2017} or a structural phase transition \cite{Kaneko2013}. This implies that the condensate could be coupled to other excitations such as plasmons or phonons. These then provide a more direct probe to study the EI collective modes by hybridizing with them.
	In particular, we consider the case where CMs hybridize with phonons. \revised{This case was explored extensively in a recent study \cite{Murakami2020Phonons}, where it was shown that hybridization with phonons would produce a CM signature in the system optical conductivity. Retaining our focus on transport properties, we note that} the hybridized mode would gain a group velocity on the electronic scale, allowing lattice distortions to propagate at velocities much faster than the normal speed of sound, and this could be measured in response to mechanical perturbations. 
	Beyond this, the robustness of the CMs to disorder scattering, compounded by their high velocity, could  impart a long ballistic propagation length to the hybridized mode. The phonon-CM hybrid may then be more mobile then the charge carriers in the material, leading to long-range lattice dynamics and correlations which do not appear in the normal phase. This could explain recent spatially-resolved pump-probe experiments on \TNS{} \cite{Andrich2020}.

	\heading{Summary}
	In this work we investigated the transport properties of collective modes in the excitonic insulator phase. We have developed a method to calculate the effects of disorder on the collective modes from a microscopic model. We have used this to demonstrate that the collective modes are robust against scattering by disorder which respects the symmetry that is spontaneously broken in the condensate. We showed that this is not a feature unique to our particular model, but is rather  a general property of the relationship between the collective modes and the condensate symmetry, by reproducing our results in a phenomenological effective field theory. We predict that collective modes could have striking consequences for the thermal conductivity, and could be probed by mechanical response. We suggest that this could explain recent spatio-temporal measurements in \TNS{}.
	
	We thank P. Andrich, H. Bretscher and A. Rao for stimulating discussions about their experimental work, and to
	D. Gole{\v{z}}, Y. Murakami, D. Bennett, O. Hart, M. McGinley, and A. Szabo for insightful remarks.
	This work was supported by EPSRC Grant No. EP/P034616/1 and by a Simons Investigator Award.
	B.~R.~ also gratefully acknowledges  the support of the Cambridge International Trust and Wolfson College, Cambridge.

	\bibliographystyle{apsrev4-1}
	\bibliography{references.bib,auxilary_refs.bib}

\begin{thebibliography}{75}
\bibitem [{par()}]{particle-hole}%
  \BibitemOpen
  \href@noop {} {}\bibinfo {note} {However, it may be shown that in the
  particle-hole symmetric case, where $J_c = J_v$, {the two sectors are
  uncoupled at any $k$}. For this reason we avoid the symmetric case, as this
  obscures the effects of mixing between the regular charge density $\hat{n}^0$
  and the other electronic degrees of freedom.}\BibitemShut {Stop}%
\end{thebibliography}
	
	%TC:ignore 

	\appendix
	%\pagebreak

	%\begin{widetext}
	
	\section{Ground State Properties}
	\label{app:GroundStateProperties}
	
	\heading{Mean-field equilibrium solution} % \label{subsec:MeanFieldGroundState}
	The ground state of the Hamiltonian \eqref{eq:Hubbard_Hamiltonian} can be found in the mean-field approximation, as worked out in Ref.~\cite{Murakami2017}. Here we briefly outline the results obtained therein, focusing on the $T=0$ limit.
	
	The electron--electron interaction is decoupled according to the prescription \cite[supplementary information (SI)]{Murakami2017}
	\begin{align} \label{InteractionMeanFieldDecoupling}
	n_{i,c} n_{i,v}  
	\!\rightarrow\!  
	\left<n_{c}\right>n_{i,v} + n_{i,c}\left<n_{v}\right>
	-\left<\phi\right>c^\d_{i,v}c^{\phantom{\dagger}}_{i,c} -\left<\phi\right>^* \! c_{i,c}^\d c^{\phantom\dagger}_{i,v}.
	\end{align}
	Here we took the equilibrium ground state to be homogeneous so that the mean fields are uniform, $\braket{\phi_i} = \braket{\phi}$, etc.
	The Hamiltonian is now quadratic in all fields, and takes the form \cite[SI, Eq.~(6)]{Murakami2017}
	\begin{equation} \label{eq:MeanFieldHamiltonian}
	H_\mathrm{el} = \frac{1}{2}\sum_k \tilde{\Psi}^\d_k \left[C_k + \vec{B}_k \cdot \vec{\sigma}\right] \tilde{\Psi}^{\phantom{\d}}_k\,
	\end{equation}
	with $\vec{\sigma}$ the vector of Pauli matrices, and $C_k,\, \vec{B}_k$ functions of the mean fields, given in Ref.~\cite[SI Eqs. (7, 11)]{Murakami2017}. 
	$\vec{B}_k$ represents a pseudo-magnetic field around which decoupled Anderson pseudospins \cite{Anderson1958}  $\frac{1}{2}\tilde{\Psi}^\d_k \vec{\sigma} \tilde{\Psi}_k$ precess. \revised{In this picture, the Falicov-Kimball model maps to an asymmetric Hubbard model \cite{Batista2002, Batista2004}.}
	
	This Hamiltonian is easily diagonalized by Bogoliubov-de Gennes \cite{FetterWalecka} modes, which hybridize the original conduction and valence states. These we denote by $\phi_{k,\pm}$ with energies $E_\pm \left(k\right) = \frac{1}{2}\left(C_k \pm \left|B_k\right|\right)$, thus implying a gapped spectrum so long as $\left|B_k\right|\neq0$.  We tune the system to be at half filling by fixing $\mu = 0$, such that all the $\phi_-$ ($\phi_+$) modes are occupied (vacant). 
	The mean-field ground state is then 
	% \begin{equation}
	$ \ket{0} = \prod_{k\in BZ} \phi^\d_{k,-} \ket{vac}.$
	% \end{equation} 
	
	The mean fields are solved by satisfying the self-consistency equations \cite[SI Eq. (10)]{Murakami2017}. 
	We choose $\braket{\phi}$ to be real without loss of generality. This choice sets $B^y_k=0$ for all $k$, so that the magnetic field lies on the XZ plane. We henceforth assume the values of the mean fields $\braket{\phi}, \braket{n_{c,v}}$ and the pseudo-magnetic field $\vec{B}_k$ (and thus $E_\pm \!\left(k\right)$) are known. 
	
	\begin{figure*}[tb]
		\centering
		\begin{subfigure}[b]{0.49\linewidth}
			\includegraphics[width=1.0\linewidth]{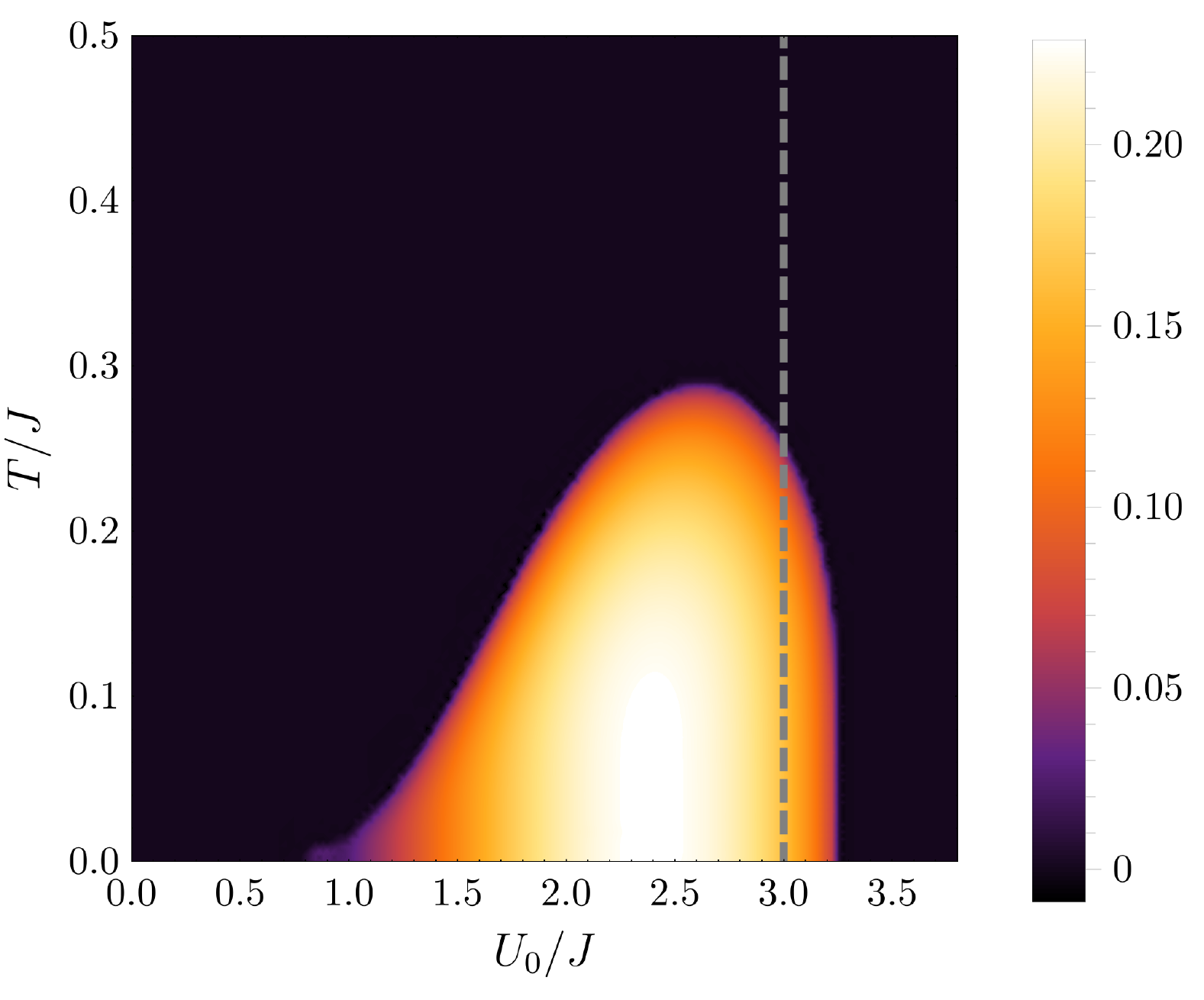}
			\caption{Contours of the order parameter $\braket{\phi}$.}
		\end{subfigure}
		\hfill
		\begin{subfigure}[b]{0.49\linewidth}
			\includegraphics[width=1.0\linewidth]{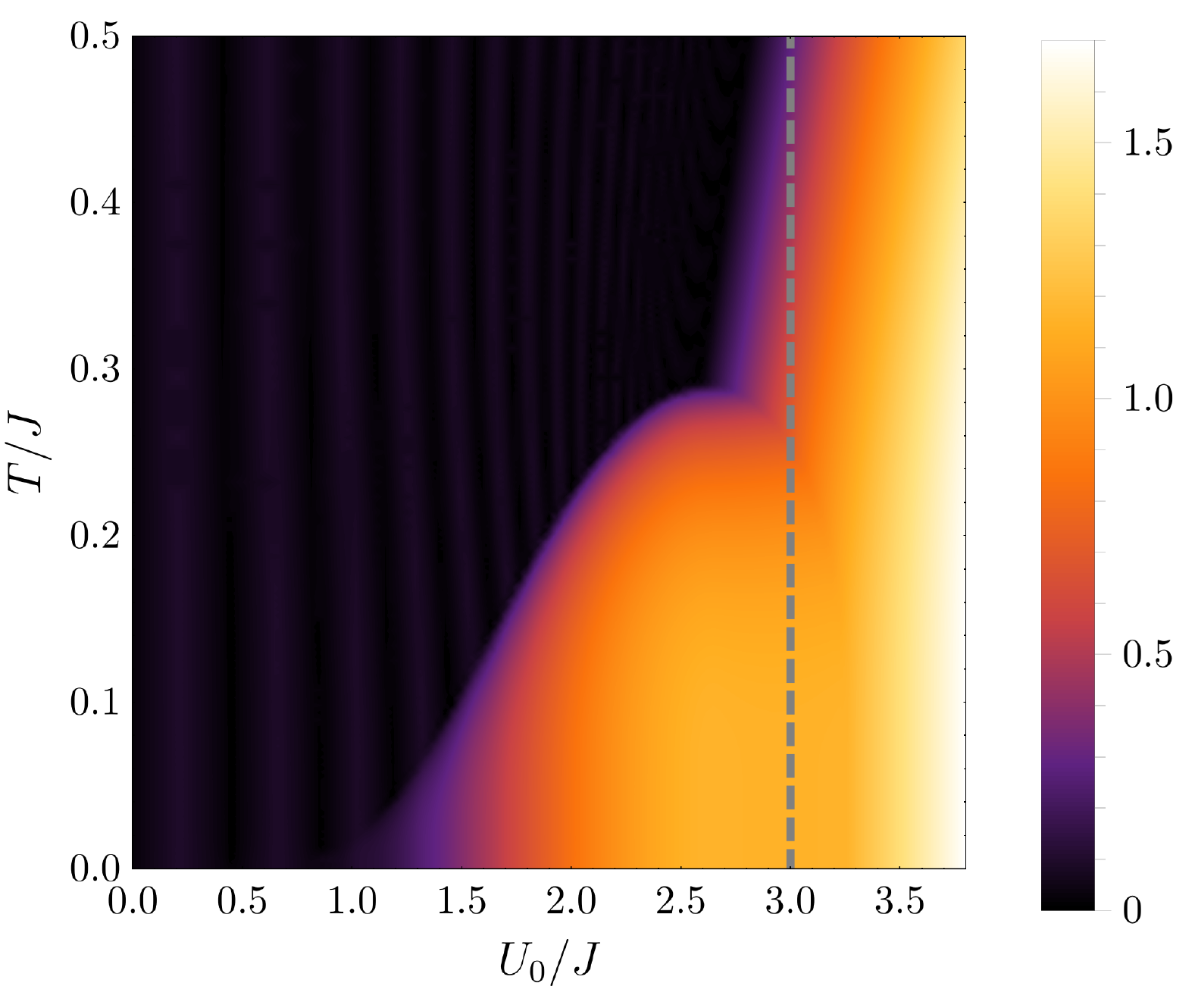}
			\caption{Contours of the band gap $\Delta$.}
		\end{subfigure}
		\caption{Calculated phase diagram of model \eqref{eq:Hubbard_Hamiltonian}, as a function of temperature $T$ and on-site repulsion $U_0$. For simplicity, the phase diagram was charted with particle-hole symmetric bands $(J_c=J_v)$. (a) Contour map of the order parameter $\braket{\phi}$, showing that the EI lies at the interface of the metal-insulator transition. (b): Contour plot of the band gap $\Delta$. The gapless region represents the normal semimetallic phase, and the uncondensed (i.e. dark in the left pane) gapped region represents the normal semiconducting phase. \revised{The metal--insulator phase boundary line bisects the condensed phase along the path of steepest ascent. This marks the crossover between the BCS and BEC regimes, lying on the left and right sides of this line, respectively. The dashed vertical line marks the value of $U_0$ used in the main text (with $T=0$), showing that the model is on the semiconducting side and hence in the BEC regime.}}
		\label{fig:HubbardModelPhaseDiagram}
	\end{figure*}
	
	With this procedure we can map the phase diagram of the model versus temperature and the interaction strength $U_0$, which tunes the band gap. This phase diagram is shown in Fig.~\ref{fig:HubbardModelPhaseDiagram} by the dependence of the order parameter $\braket{\phi}$ and the band gap $\Delta$ on these parameters. The value of $U_0$ used in the main text is indicated by the dashed line. \revised{It lies on the semiconducting (i.e. gapped) side of the normal phase diagram, indicating that the parameters used in the main text place the excitonic condensate in the BEC regime.}

	\heading{Feynman rules}
	We now list the Feynman diagrammatic rules used to evaluate the expansion \eqref{eq:ChiWickTheoremConnected}. The electron time-ordered propagator is, 
	\begin{align} \label{eq:SingleParticleTimeOrderedBareGTimeDomain}
	& \begin{tikzpicture}[baseline={([yshift=-.5ex]current bounding box.center)}]
	\draw [-{Stealth[sep=-5pt]}] (-1,0) node[anchor=east] {$\tilde{c}^\d_\beta \left(0\right)$}
	-- (0,0) node[anchor=south] {$k$} ;
	\draw  (-1,0) -- (1,0)  node[anchor=west] {$\tilde{c}_\alpha \left(t\right)$};
	\end{tikzpicture} =
	iG^0_{\alpha\beta}\left(k,t\right)  \tcbr =
	\Theta\left(t\right)iG_{\alpha\beta}^>\left(k,t\right)+
	\Theta\left(-t\right)iG_{\alpha\beta}^<\left(k,t\right)\,.
	\end{align}
	Here $G^{\{>,<\}}$ are the electronic one-particle Green's functions, defined by 
	$ G^>_{k,\alpha\beta}\left(t\right)= -i\braket{0|\tilde{c}^{\phantom{\d}}_{k\alpha}\left(t\right) \tilde{c}^\d_{k\beta}\left(0\right)|0}$ and
	$G^<_{k,\alpha\beta}\left(t\right)= +i\braket{0|\tilde{c}^\d_{k\beta}\left(0\right) \tilde{c}^{\phantom{\d}}_{k\alpha}\left(t\right)|0}$,
	where $\alpha,\beta$ are electron band indices (conduction or valence). Defining the matrices $M_\pm \left(k\right) = \mathds{1} \pm \vec{B}_k\cdot \vec{\sigma}$, they evaluate to \cite[SI Eq. (12), at $T=0$]{Murakami2017}
	\begin{align}
	& G^>_k (t)=  -\frac{i}{2}  e^{-iE_+\!(k)t}M_+(k), \, %\qquad \tcbr
	G^<_k (t) =  \frac{i}{2}  e^{-iE_-\!(k)t}M_-(k).
	\end{align}
	In the frequency domain the time-ordered propagator then takes the form
	\begin{equation}\label{eq:SingleParticleTimeOrderedBareG}
	G^0_{\alpha\beta}\left(k,\omega\right)=\frac{1}{2}\left[
	\frac{M_+\left(k\right)}{\omega^+-E_+\left(k\right)} + 
	\frac{M_-\left(k\right)}{\omega^--E_-\left(k\right)}
	\right]_{\alpha\beta}\,.
	\end{equation}
	
	\def \vertexsize {0.5} 
	
	{To facilitate the electron--electron interaction, we rewrite the interaction term as 
		\begin{align}
		\hat{U} = \sum_i U_0 n_{i,c} n_{i,v}  	= \hat{U}_\mathrm{MF} + \left(\hat{U}  - \hat{U}_\mathrm{MF}\right)	  
		\end{align}
		where $\hat{U}_\mathrm{MF}$ is the operator obtained by employing prescription \eqref{InteractionMeanFieldDecoupling}. 
		By absorbing the first term into the zeroth-order Hamiltonian, it will define the same mean-field ground state as before. The second term defines the  electron--electron 4-vertex interaction.
		Note that any contraction of this vertex with itself vanishes by construction. Remembering to remove these diagrams manually, we can take the simplified vertex 
		\begin{align}
		\hat{U} & - \hat{U}_\mathrm{MF} \rightarrow \hat{U} = 
		\begin{tikzpicture}[baseline={([yshift=-.5ex]current bounding box.center)}]
		\begin{feynman}
		\vertex (LU) at (-\vertexsize,\vertexsize) ;
		\vertex (LD) at (-\vertexsize,-\vertexsize) ;
		\vertex (RD) at (\vertexsize,-\vertexsize) ;
		\vertex (RU) at (\vertexsize, \vertexsize) ;
		\vertex (C) at (0, 0) ;
		\diagram* {
			(RU) -- [fermion] (C) -- [fermion] (RD),
			(LD) -- [fermion] (C) -- [fermion] (LU)
		};
		\end{feynman}
		\end{tikzpicture}
		= \sum_i U_0 n_{i,c} n_{i,v} \tcbr
		= \frac{U_0}{N}\sum_{\lbrace k_i \rbrace} \delta^\text{BZ} \left(k_1 + k_3 - k_2 - k_4 \right)
		\tilde{c}^\d_{c,k_1}\tilde{c}_{c,k_2}\tilde{c}^\d_{v, k_3} \tilde{c}_{v, k_4}\,.
		\end{align}
	}
	
	Finally, we project this vertex onto the sector of exciton RPA diagrams. Recall that it has no self-contractions. Thus, there are four ways to contract one such vertex with the external vertices which define $\chi$ (cf. Eq.~\eqref{eq:ChiWickTheoremConnected}). Summing over these combinations (while suppressing momentum indices for brevity),
	\begin{align}
	&\braket{\mathcal{T}\lbrace  \hat{U}\hat{n}^\mu\hat{n}^\nu  \rbrace}_0 	= \chi^0_{\mu\rho} \left[\frac{1}{2N}U_0\left( 2 \delta_{\rho0}\delta_{\sigma0} -\delta_{\rho \sigma} \right)\right] \chi^0_{\sigma\nu} \,.
	\end{align}
	This implies that in the exciton picture, the interaction vertex takes the form 
	\begin{equation}
	\hat{U}_{\mu\nu} =
	\frac{1}{2N} U_0 
	\delta \left(k_1 - k_2\right) 
	\mathrm{diag}\left(1, -1, -1, -1\right) 
	\end{equation}
	where $k_{1,2}$ are the net momenta of the incoming and outgoing excitons. 
	The vertex is proportional to $N^{-1}$; yet, a single bubble diagram in $k$-space is equal to $iN\chi$, so factors of $N$ cancel and we thus suppress them in the main text. 
	
	\heading{Zeroth-order linear response}
	Finally, we compute $\chi^0$. It is found by the loop integration over the bubble diagram 
	\begin{align} \label{eq:Chi0DiagrammaticEvaluation}
	\chi_{\mu\nu}^0 & \!\left(k,\omega\right)  
	= -i \times \sigma^\mu \, \bubble \, \sigma^\nu 
	\tcbr
	= \frac{1}{N} \sum_p \int \frac{d \omega'}{2 \pi i} 
	\tr \left[G^0_{p} \! \left(\omega'\right) \sigma^\mu G^0_{p+k} \! \left(\omega' + \omega \right) \sigma^\nu \right]\,.
	\end{align}
	Recall this must still be retarded.
	Substituting Eq.~\eqref{eq:SingleParticleTimeOrderedBareG} into Eq.~\eqref{eq:Chi0DiagrammaticEvaluation}, performing the convolution  and retarding (transcribing $\omega^- \rightarrow \omega^+$), we find \cite{Murakami2020, Murakami2020Phonons} the Lehmann representation \cite{FetterWalecka} of $\chi^0$
	\begin{align} \label{eq:BareChi0Lehmann}
	\chi^0_{\mu\nu}(k, \omega ) = \frac{1}{4N} &  \sum_{p}  \Bigg[ \frac{A_{+-}^{\mu\nu}(p, p-k)}{\omega^+ - \left(E_+(p)-E_-(p-k)\right)}  \tcbr
	- \frac{A_{-+}^{\mu\nu}(p, p-k)}{\omega^+ - \left(E_-(p)-E_+(p-k)\right)} \Bigg]\,,
	\end{align}
	with the shorthand notation $A_{\alpha\beta}^{\mu\nu}\left(p,q\right) = \tr \left[\sigma^\mu M_\alpha \left(p\right) \sigma^\nu M_{\beta}\left(q\right)\right]$.
	
	From Eq.~\eqref{eq:BareChi0Lehmann} we express the static bare susceptibility, which controls the disorder screening (cf. Sec.~\ref{sec:Disorder})
	\begin{equation}
	\chi^0_{\mu\nu} \! \left(k, \omega = 0\right) 
	= -\frac{1}{2N} \sum_p \frac{\mathrm{Re}~A^{\mu\nu}_{+-}\left(p,p-k\right)}{E_+\left(p\right) - E_-\left(p-k\right)}\,.
	\end{equation}
	From this it follows that $\chi^0_{y\nu}=\chi^0_{\mu y} = 0$, so that $\chi^0_{yy}$ does not mix with the other components in the RPA solution \eqref{eq:chiRPASolution}. Therefore, the screened $\lambda = y$ disorder channel remains uncorrelated with the others.
	
	%%%%%%%%%%%%%%%%%%%%%%%%%%%%%%%%%%%%%%%%%%%%%%%%%%%%%%%%%%%%%%%%%
	%%%%
	%%%%			APPENDIX BREAK
	%%%%
	%%%%%%%%%%%%%%%%%%%%%%%%%%%%%%%%%%%%%%%%%%%%%%%%%%%%%%%%%%%%%%%%%

	\section{Quasiparticle Self-Energy} \label{app:QuasiparticleSelfEnergy}
	Here we evaluate the quasipartice self-energy in Eq.~\eqref{eq:SingleParticleSelfEnergyRate} to find their scattering rates. We will use this as an instructive example to demonstrate the same procedure we will employ for the exciton self-energy in Appendix \ref{app:ExcitonSelfEnergyDiagrammatics}. 
	
	As we require the imaginary part of a diagonal element of the self-energy, we are interested only in the anti-Hermitian part of $\Sigma$. We thus write
	Eq.~\eqref{eq:SingleParticleSelfEnergyDiagram}
	\begin{align} 
	& \Sigma\left(k,\omega\right) 
	= \frac{1}{N} \!\! \sum_{\lambda, \lambda',q} \!\! \tilde{\mathbb{K}}_q^{\lambda\lambda'} \sigma^{\lambda} G_{k + q}^0 \!\left(\omega\right) \sigma^{\lambda'} 
	\tcbr
	\Rightarrow 
	\frac{1}{N} \!\! \sum_{\lambda, \lambda',q} \!\! \tilde{\mathbb{K}}_q^{\lambda\lambda'} \sigma^{\lambda} 
	\frac{1}{2}\left[G_{k + q}^0 \!\left(\omega\right) - \left(G_{k + q}^0\right)^\d \!\left(\omega\right)\right] \sigma^{\lambda'}\,,
	\end{align}
	where we have used that $\tilde{\mathbb{K}}_q$ is a symmetric matrix. Substituting Eq~\eqref{eq:SingleParticleTimeOrderedBareG}, we find 
	\begin{align*}
	& \frac{1}{2}\left[G_{p}^0 \!\left(\omega\right) - \left(G_{p}^0\right)^\d \!\left(\omega\right)\right] \tcbr
	= - \frac{i}{2} \pi \left[ M_+(p) \delta(\omega - E_+ (p))  
	-  M_-(p) \delta(\omega - E_- (p)) \right]\,.
	\end{align*}
	We are interested in positive energy states at $\omega = E_+\left(k\right)$, so henceforth suppress the second term. 
	
	Next we take the continuum limit to obtain the integration measure
	\begin{align}
	\frac{1}{N} \!\! \sum_q & \delta \! \left(E_+(k) - E_+(k\!+\!q)\right) 
	= \!\!\int\!\! \frac{a dq}{2 \pi } \!\! \left[ \! \frac{\delta\left(q\right)}{|E'_+(k)|} \!+\! \frac{\delta\left(q + 2k\right)}{|E'_+(-k)|} \!\right] 
	\tcbr= \frac{1}{2} \int dq \, g\left(E_+(k)\right) \left[\delta\left(q\right) + \delta \left(q+2k\right)\right]\,
	\end{align}
	with $g\left(E_+(k)\right) = 2\times \frac{a}{2\pi}\left|\frac{dk}{dE_+\left(k\right)}\right|$ the single-particle density of states, the  factor of two rising from the $\pm k $ degeneracy. 
	
	Combining results, we have 
	\begin{align*}
	-&\mathrm{Im}~[\Sigma  \left(k,E_+(k)\right)] \tcbr = 
	\frac{\pi}{2}  g\left(E_+(k)\right) \sum_{\lambda \lambda'} 
	[\tilde{\mathbb{K}}^{\lambda \lambda'}_{0} + \tilde{\mathbb{K}}^{\lambda \lambda'}_{-2k}] \frac{1}{2} [\sigma^\lambda  M_+(k) \sigma^{\lambda'}]\,.
	\end{align*}
	To obtain the scattering rate, we must rotate and extract the diagonal element per Eq.~\eqref{eq:SingleParticleSelfEnergyRate}. Noting that $M_+\left(k\right) = 2 R^{\phantom{\d}}_k \mathrm{diag}\left(1,0\right) R^\d_k$, we have 
	\begin{equation}
	\frac{1}{2} [R^\d_k \sigma^\lambda  M_+\left(k\right) \sigma^{\lambda'} R^{\phantom{\d}}_k]_{11} 
	= \hat{B}_k^\lambda \hat{B}_k^{\lambda'}\,.
	\end{equation}
	
	We thus obtain the scattering rate
	\begin{equation} \label{eq:SingleParticleDecayRateFull}
	\Gamma^{\mathrm{QP}}_k =  \frac{1}{2} \pi g\left(E_+\left(k\right)\right) \sum_{\lambda \lambda'} \hat{B}_k^\lambda \left[\tilde{\mathbb{K}}^{\lambda \lambda'}_{0} + \tilde{\mathbb{K}}^{\lambda \lambda'}_{-2k}\right]  \hat{B}_k^{\lambda'} 
	\,.
	\end{equation}
	The two contributions in the bracket correspond to forwards and backwards scattering, respectively. However, forwards scattering does not affect transport, and we therefore discard this term in the results shown in Sec.~\ref{sec:Disorder}.

	%%%%%%%%%%%%%%%%%%%%%%%%%%%%%%%%%%%%%%%%%%%%%%%%%%%%%%%%%%%%%%%%%
	%%%%
	%%%%			APPENDIX BREAK
	%%%%
	%%%%%%%%%%%%%%%%%%%%%%%%%%%%%%%%%%%%%%%%%%%%%%%%%%%%%%%%%%%%%%%%%

	\section{Evaluation of Exciton Self-Energy Diagrams} \label{app:ExcitonSelfEnergyDiagrammatics}
	In this appendix we evaluate the exciton self-energy diagrams up to second order in the disorder potential. 
	All RPA-type second order diagrams are enumerated in Eq.~\eqref{eq:DisorderDiagrams}, and fall into two categories: those where the disorder vertices contract within the same bubble (first two diagrams), and those where they contract between different bubbles (last two). For reasons which will become apparent below, we call the first type subresonant diagrams, and the second scattering diagrams.

	\def \bubblesize {4.0}
	
	\begin{figure*}[tb]
		\centering
		\begin{equation*}
		\bubblePurse \quad = \quad
		% Purse iagram Up
		\begin{tikzpicture}[baseline={([yshift=-.5ex]m.base)}]
		\begin{feynman}
		\vertex (m) at (0,0) ;
		\vertex (n) at (+\bubblesize,0) ;
		% The displacements relative to the center of the arc are
		% dx = R * sin(\alpha) = \bubblesize * sin(\alpha) / \sqrt(2)
		% dy = dx * tan(\alpha/2)
		% for simplicity we pick here \alpha = \pi/6.
		\vertex (u1) at (\bubblesize*0.5 - \bubblesize*0.353553, \bubblesize*0.414213*0.5 - \bubblesize*0.0947343);
		\vertex (u2) at (\bubblesize*0.5 + \bubblesize*0.353553, \bubblesize*0.414213*0.5 - \bubblesize*0.0947343);
		
		\vertex[below left=0.05cm of u2] (u2l) {\(V^{\lambda_1}_q\sigma^{\lambda_1}_{\rho\sigma}\)};
		\vertex[below right=0.05cm of u1] (u1l) {\(V^{\lambda_2}_{-q}\sigma^{\lambda_2}_{\eta\theta}\)};
		\vertex[below left=0.05cm of m] (ml) {\(\sigma^\mu_{\alpha\beta}\)};
		\vertex[below right=0.05cm of n] (nl) {\(\sigma^\nu_{\gamma\delta}\)};
		
		\diagram* {
			(m) -- [fermion, out = 45, in = 210, edge label=\({p,\omega'}\)] (u1) -- [fermion, out = 30, in = 150, edge label=\({p-q,\omega'}\)] (u2) -- [fermion, out = -30, in = 135, edge label=\({p,\omega'}\)] (n) -- [fermion, out = -135, in = -45, edge label=\({p+k,\omega+\omega'}\)] (m),
			(u1) -- [boson, half left, very thin, edge label'=\(q\)] (u2) ,
			(u1) -- [charged boson, red, thick, half left] (u2),
		};
		\end{feynman}
		\end{tikzpicture}
		\quad + \quad 
		\begin{tikzpicture}[baseline={([yshift=-.5ex]m.base)}]
		\begin{feynman}
		\vertex (m) at (0,0) ;
		\vertex (n) at (+\bubblesize,0) ;
		% The displacements relative to the center of the arc are
		% dx = R * sin(\alpha) = \bubblesize * sin(\alpha) / \sqrt(2)
		% dy = dx * tan(\alpha/2)
		% for simplicity we pick here \alpha = \pi/6.
		\vertex (u1) at (\bubblesize*0.5 + \bubblesize*0.353553, -\bubblesize*0.414213*0.5 + \bubblesize*0.0947343);
		\vertex (u2) at (\bubblesize*0.5 - \bubblesize*0.353553, -\bubblesize*0.414213*0.5 + \bubblesize*0.0947343);
		
		\vertex[above right=0.05cm of u2] (u2l) {\(V^{\lambda_1}_q\sigma^{\lambda_1}_{\rho\sigma}\)};
		\vertex[above left=0.05cm of u1] (u1l) {\(V^{\lambda_2}_{-q}\sigma^{\lambda_2}_{\eta\theta}\)};
		\vertex[above left=0.05cm of m] (ml) {\(\sigma^\mu_{\alpha\beta}\)};
		\vertex[above right=0.05cm of n] (nl) {\(\sigma^\nu_{\gamma\delta}\)};
		
		\diagram* {
			(n) -- [fermion, out = -135, in = 30, edge label=\({p,\omega'}\)] (u1) -- [fermion, out = 210, in = -30, edge label=\({p-q,\omega'}\)] (u2) -- [fermion, out = 150, in = -45, edge label=\({p,\omega'}\)] (m) -- [fermion, out = 45, in = 135, edge label=\({p-k,\omega-\omega'}\)] (n),
			(u1) -- [boson, half left, very thin, edge label'=\(q\)] (u2) ,
			(u1) -- [charged boson, red, thick, half left] (u2),
		};
		\end{feynman}
		\end{tikzpicture}
		\end{equation*}
		\caption{The exciton ``purse'' diagram.}
		\label{fig:ExcitonPurseDiagram}
	\end{figure*}
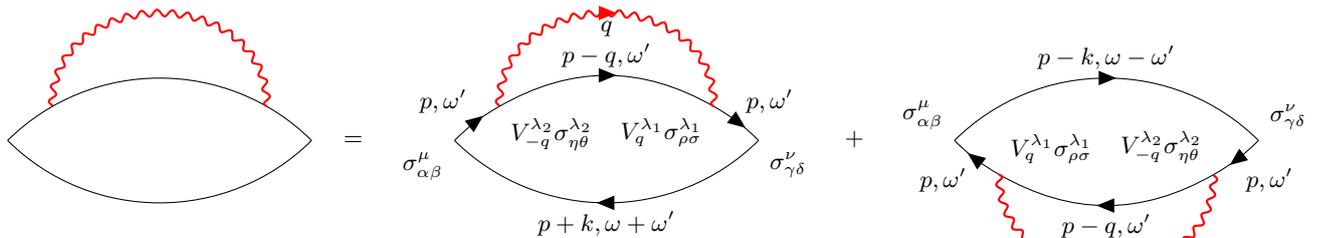
	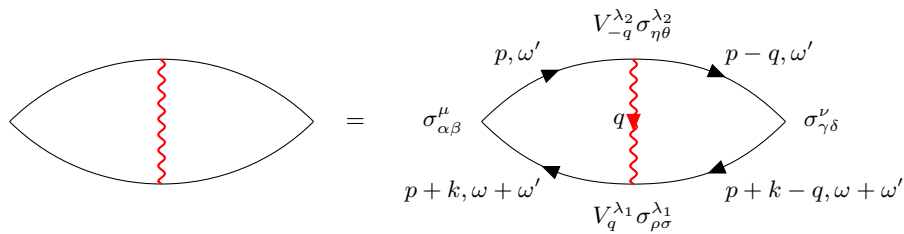
\begin{figure*}[tb]
		\centering
		\begin{equation*}
		\bubbleClam \quad = \quad
		% Clam Diagram
		\begin{tikzpicture}[baseline={([yshift=-.5ex]m.base)}]
		\begin{feynman}
		\vertex (m) at (0,0) ;
		\vertex (n) at (+\bubblesize,0) ;
		\vertex(u1) at (\bubblesize*0.5,  \bubblesize*0.414213*0.5);
		\vertex(u2) at (\bubblesize*0.5, -\bubblesize*0.414213*0.5);
		
		\vertex[below=0.125cm of u2] (u2l) {\(V^{\lambda_1}_q\sigma^{\lambda_1}_{\rho\sigma}\)};
		\vertex[above=0.125cm of u1] (u1l) {\(V^{\lambda_2}_{-q}\sigma^{\lambda_2}_{\eta\theta}\)};
		\vertex[left=0.125cm of m] (ml) {\(\sigma^\mu_{\alpha\beta}\)};
		\vertex[right=0.125cm of n] (nl) {\(\sigma^\nu_{\gamma\delta}\)};
		
		\diagram* {
			(m) -- [out = 45, in = 180, edge label=\({p,\omega'}\), fermion] (u1) -- [fermion, out = 0, in = 135, edge label=\({p-q,\omega'}\)] (n) -- [fermion, out = -135, in = 0, edge label=\({p+k-q,\omega+\omega'}\)] (u2) -- [fermion, out = 180, in = -45, edge label=\({p+k,\omega+\omega'}\)] (m),
			(u1) -- [boson, very thin, edge label'=\(q\)] (u2) ,
			(u1) -- [charged boson, red, thick] (u2)
		};
		\end{feynman}
		\end{tikzpicture}	
		\end{equation*}
		\caption{The exciton ``clam'' diagram.}
		\label{fig:ExcitonClamDiagram}
	\end{figure*}
	
	\heading{Subresonant diagrams}
	The subresonant diagrams, which we will call ``purse'' and ``clam'' diagrams, are shown in Figs.~\ref{fig:ExcitonPurseDiagram} and \ref{fig:ExcitonClamDiagram}, respectively. 
	Since these diagrams contain an internal contraction of the disorder potential, they can be ensemble-averaged immediately. The diagrams then evaluate to
	
	\begin{widetext}
		\begin{equation} \label{eq:ExcitonPurseMatrixElement}
		\Sigma^\mathrm{purse}_{\mu\nu}  \! \left( k, \omega \right) = -\frac{i}{N^2}
		\!\! \sum_{\substack{ \lambda_1, \lambda_2 \\ p,q }} \!\!
		\tilde{\mathbb{K}}^{\lambda_1 \lambda_2}_q \! \! \int \frac{d\omega '}{2 \pi } \tr \lbrace
		G^0_{p}\left(\omega'\right) \sigma^{\lambda_1} G^0_{p-q} \left(\omega'\right) \sigma^{\lambda_2} G^0_p \left(\omega'\right) 
		\left[ \sigma^\mu G^0_{p+k}\left(\omega' + \omega\right) \sigma^\nu \!+ \sigma^\nu G^0_{p-k}\left(\omega'-\omega\right) \sigma^\mu \right]
		\rbrace,
		\end{equation}
		which shows the transposition symmetry $\mu \leftrightarrow \nu,\, \omega\rightarrow -\omega,\, k \rightarrow -k$, and
		\begin{equation} \label{eq:ExcitonClamMatrixElement}
		\Sigma^\mathrm{clam}_{\mu\nu} \! \left( k, \omega \right) = -\frac{i}{N^2}
		\sum_{\substack{ \lambda_1, \lambda_2 \\ p,q }} 
		\tilde{\mathbb{K}}^{\lambda_1 \lambda_2}_q \int \frac{d\omega '}{2 \pi } \tr \lbrace
		G^0_{p-q}\left(\omega'\right) \sigma^{\lambda_2} G^0_{p} \left(\omega'\right) \sigma^{\mu} G^0_{p+k}\left(\omega'+\omega\right) \sigma^{\lambda_1} G^0_{p+k-q}\left(\omega' + \omega\right) \sigma^\nu
		\rbrace\,,
		\end{equation}		\end{widetext}
	which does not. 
	
	We call these diagrams subresonant as they cannot be interpreted as representing on-shell scattering of a collective mode from an initial to a final state. The identification of self-energy diagrams with the FGR requires an internal virtual state, which would be represented by $\chi^\text{RPA}$. Indeed, Eqs.~\eqref{eq:ExcitonPurseMatrixElement} and \eqref{eq:ExcitonClamMatrixElement} contain products of poles above and below the real $\omega'$ line (cf. Eq.~\eqref{eq:SingleParticleTimeOrderedBareG}), and their convolutions can only mix poles from opposite sides. Thus, both matrix elements above only have poles at frequencies $\pm \left(E_+\left(k\right) - E_-\left(k'\right)\right)$ which lie in the two-particle continuum. No poles exist at lower energies, so the diagrams are not resonant at the collective mode energies, so long as they are below the continuum edge. To put it simply, since $\chi^\text{RPA}$ does not appear in these diagrams, they do not ``know'' about the collective modes.
	
	Since they are not resonant, we do not expect them to contribute to the collective mode scattering rates. 		We can prove this by showing that these diagrams add Hermitian terms to the self-energy, which thus have purely real diagonal elements in any basis.
	
	This is straightforward to show for the purse diagram:  In the time domain $\chi\left(i,t\right)$ is real, and thus $\chi\left(-k,-\omega \right) = \chi\left(k, \omega \right)^*$.  We now note that $\Sigma$ itself is the leading-order correction to $\chi$ in Eq.~\eqref{eq:DisoerderResummation}, and must thus respect this property. It is tempting to merge this fact with the transposition symmetry of  $\Sigma^\text{purse}\left(k,\omega\right)$ and conclude that it is Hermitian for all $k,\omega$. However, this would imply that even in the continuum this diagram has no associated scattering rate. The crucial point to note is that $\Sigma^\text{purse}$ shown above is \emph{time ordered}. 		However, the time-ordered and retarded $\Sigma\left(k,\omega\right)$, at fixed $k$, differ only by the infinitesimal shift of their poles, i.e. where their branch cuts run with respect to the real line. At frequencies below the continuum edge they are pole-free, and thus they coincide. Ergo, below the continuum the relation $\Sigma\left(-k,-\omega \right) = \Sigma\left(k, \omega \right)^*$ applies, and $\Sigma^\text{purse}$ is indeed Hermitian.
	
	For the clam diagram, we note  that $\left(G^0_k\right)^T = G^0_k$, and thus $\left(G^0_k\right)^\d = G^0_k \big|_{\eta\rightarrow -\eta}$. We then write 		\begin{widetext}
		\begin{equation*}
		\Sigma^\text{clam}_{\mu\nu} \! \left( k, \omega \right)= 
		\frac{1}{2N^2} \! \! \sum_{\substack{ \lambda_1, \lambda_2 \\ p,q }} \!\!
		\tilde{\mathbb{K}}^{\lambda_1 \lambda_2}_q \lim_{\eta\rightarrow 0} \int \! \frac{d\omega'}{2 \pi i} \tr \lbrace
		G^0_{p-q}\left(\omega'\right) \sigma^{\lambda_2} G^0_{p} \left(\omega'\right) \sigma^{\mu} G^0_{p+k}\left(\omega'+\omega\right) \sigma^{\lambda_1} G^0_{p+k-q}\left(\omega' + \omega\right) \sigma^\nu
		+\mathrm{h.c.}\big|_{\eta \rightarrow -\eta}
		\rbrace,
		\end{equation*}
	\end{widetext}
	where $\mathrm{h.c.}$ refers to conjugate transposition of the indices $\mu,\nu$, rather then the matrix which is traced. Recall this convolution produces poles only in the continuum. For $\omega$ below the continuum, we do not encounter any branch cuts and the limit of $\eta\rightarrow0$ is symmetrical.  However, the operation of flipping $\eta$ does not commute with contour integration; since all poles are mirrored, in the second term we need to close the contour around the opposite half-plane. Therefore
	\begin{equation*}
	\lim_{\eta \rightarrow 0} \int \frac{d\omega'}{2 \pi i} \left(\bullet \right)^\d_{\eta \rightarrow -\eta} 
	= - \int \frac{d\omega'}{2 \pi i} \left(\bullet\right)^\d
	= \left( \int \frac{d\omega'}{2 \pi i} \, \bullet \right)^\d\,,
	\end{equation*}
	where it is the first equality that is valid only below the continuum. We thus find $\Sigma^\text{clam} = \frac{1}{2} \Sigma^\mathrm{clam} + \mathrm{h.c.}$, proving that it is Hermitian.

\begin{figure*}[tb]
		\centering
		\begin{equation*}
		\mathcal{V}^\lambda_{\mu\nu}\left(k_1,k_2,\omega\right)
		\quad = \quad
		\begin{tikzpicture}[baseline={([yshift=-.5ex]m.base)}]
		\begin{feynman}
		\vertex (m) at (0,0) ;
		\vertex (m0) at (-0.25*\bubblesize, 0) {\(k_1,\omega\)};
		\vertex (n) at (+\bubblesize,0) ;
		\vertex (n0) at (1.25*\bubblesize, 0) {\(k_2,\omega\)};
		\vertex (u) at (\bubblesize*0.5,\bubblesize*1.414213*0.5 - \bubblesize*0.5);
		\vertex (uu) at (\bubblesize*0.5, \bubblesize*0.5);% {\(q = k_1-k_2\)};
		
		\vertex[below=0.05cm of u] (ul) {\(V^{\lambda}_{q}\sigma^{\lambda}_{\rho\sigma}\)};
		\vertex[right=0.25cm of m] (ml) {\(\sigma^\mu_{\alpha\beta}\)};
		\vertex[left=0.25cm of n] (nl) {\(\sigma^\nu_{\gamma\delta}\)};
		
		\diagram* {
			%	(m) -- [fermion, quarter left]  (n) -- [fermion, quarter left] (m) -- [boson] (u),
			(m) -- [fermion, out = 45, in = 180, edge label=\({p,\omega'}\)] (u) -- [fermion, out = 0, in = 135, edge label=\({p+q,\omega'}\)] (n) -- [fermion, out = -135, in = -45, edge label=\({p+k_1,\omega+\omega'}\)] (m),
			(u) -- [boson, very thin, edge label'=\({q=k_1-k_2}\)] (uu),
			(u) -- [anti charged boson, red, thick] (uu),
			(m) -- [fermion, opacity = 0.5] (m0),
			(n0) -- [fermion, opacity = 0.5] (n)
		};
		\end{feynman}
		
		\end{tikzpicture}
		\quad + \quad 
		%
		% Vertex Diagram Down
		\begin{tikzpicture}[baseline={([yshift=-.5ex]m.base)}]
		\begin{feynman}
		\vertex (m) at (0,0) ;
		\vertex (m0) at (-0.25*\bubblesize, 0) {\(k_1,\omega\)};
		\vertex (n) at (+\bubblesize,0) ;
		\vertex (n0) at (1.25*\bubblesize, 0) {\(k_2,\omega\)};
		\vertex (d) at (\bubblesize*0.5,-\bubblesize*1.414213*0.5 + \bubblesize*0.5);
		\vertex (dd) at (\bubblesize*0.5, -\bubblesize*0.5);% {\(q = k_1-k_2\)};
		
		\vertex[above=0.05cm of d] (dl) {\(V^{\lambda}_{q}\sigma^{\lambda}_{\rho\sigma}\)};
		\vertex[right=0.25cm of m] (ml) {\(\sigma^\mu_{\alpha\beta}\)};
		\vertex[left=0.25cm of n] (nl) {\(\sigma^\nu_{\gamma\delta}\)};
		
		\diagram* {
			%	(m) -- [fermion, quarter left]  (n) -- [fermion, quarter left] (m) -- [boson] (u),
			(m) -- [fermion, out = 45, in = 135, edge label=\({p,\omega'}\)] (n) -- [fermion, out = -135, in = 0, edge label=\({p+k_2,\omega+\omega'}\)] (d) -- [fermion, out = 180, in = -45, edge label=\({p+k_1,\omega+\omega'}\)] (m),
			(d) -- [boson, very thin, edge label'=\({q=k_1-k_2}\)] (dd),
			(d) -- [anti charged boson, red, thick] (dd),
			(m) -- [fermion, opacity = 0.5] (m0),
			(n0) -- [fermion, opacity = 0.5] (n)
		};
		\end{feynman}
		
		\end{tikzpicture}
		\end{equation*}
		\caption{Structure of the effective exciton disorder vertex diagram.}
		\label{fig:exciton_self_energy_vertex}
	\end{figure*}
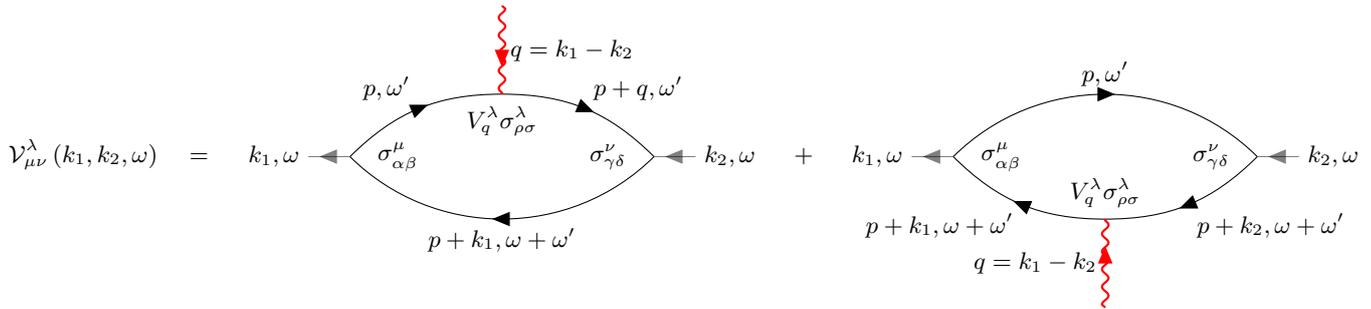

	\heading{Effective exciton disorder vertex} 
	To calculate the actual scattering diagrams, we first need to evaluate the subelement of a single bubble coupled to the disorder potential. The potential could couple to either electron or hole propagators, and we must sum over both options. This single bubble will become an effective disorder vertex which attaches on either side to $\chi$, which serves as the exciton effective propagator. An annotated sketch of this diagram is presented in Fig.~\ref{fig:exciton_self_energy_vertex}.
	Unlike the underlying potential $V_q$ which depends only on the momentum transfer $q$, this vertex will depend on both incoming and outgoing momenta, as well the energy. It evaluates to
	\begin{align}
	\mathcal{V}^\lambda_{\mu\nu}&(k_1,k_2,\omega)  = -{\tilde{\mathbb{V}}^\lambda_{q}} \frac{1}{N} \sum_p \int \frac{d\omega '}{2 \pi i} \tr \lbrace
	G^0_{p+q}(\omega') \sigma^\lambda	G^0_p (\omega') \tcbr \times
	\left[ \sigma^\mu G^0_{p+k_1}(\omega' + \omega) \sigma^\nu + \sigma^\nu G^0_{p-k_2}(\omega'-\omega) \sigma^\mu \right]
	\rbrace\,,
	\end{align}
	with $q = k_1 - k_2$. The integration over $\omega'$ may be performed analytically by plugging in the propagator \eqref{eq:SingleParticleTimeOrderedBareG}, yielding a Lehmann-type representation akin to Eq.~\eqref{eq:BareChi0Lehmann}, but this gives no additional insight. 
	
	The two terms inside the trace correspond to the sum over diagrams in Fig.~\ref{fig:exciton_self_energy_vertex}, and impart this vertex with a combined transposition symmetry $\mu \leftrightarrow \nu,\, \omega\rightarrow -\omega,\, k_1 \leftrightarrow -k_2$. Note that the last transformation flips the center-of-mass momentum $k_1 + k_2 \rightarrow -\left(k_1 + k_2\right)$, but leaves $q$ unchanged. 
	{By the same argument invoked for the subresonant diagrams, below the continuum the effective vertex is regular and thus the time-ordered and retarded $\mathcal{V}^\lambda$ coincide. Therefore, $\mathcal{V}^\lambda(-k_1,-k_2, -\omega) = \left(\mathcal{V}^\lambda (k_1,k_2,\omega)\right)^*$. Combining both symmetries, we find $\left(\mathcal{V}^\lambda(k_1,k_2,\omega)\right)^\d = \mathcal{V}^\lambda(k_2,k_1,\omega)$, i.e. it is Hermitian up to an interchange of momenta.}
	
	{This effective vertex appears by itself in the expansion \eqref{eq:ChiWickTheoremConnected}, but vanishes upon disorder averaging as it is first order in $\mathds{V}^\lambda$.} However, its contractions with itself enter the self-energy, where they are ``glued together'' by the effective vertex $U^\text{eff}$ and an arbitrary number of susceptibility bubbles, i.e. $\chi^\text{RPA}$, as shown in Eq.~\eqref{eq:DisorderDiagrams}. This is written 
	\begin{align}\label{eq:ExcitonSelfEnergyDiagram}
	\Sigma^\text{scat.} & (k, \omega )  		= \frac{1}{N} \sum_{k', \lambda_{1,2}} \Big\langle
	\mathcal{V}^{\lambda_1} (k, k', \omega) * U^\text{eff} * \tcbr
	*\mathcal{G}^\mathrm{RPA}(k',\omega) *
	U^\text{eff} *
	\mathcal{V}^{\lambda_2} (k', k, \omega) \Big\rangle		\,. 
	\end{align}
	
	We note the similarity between this self-energy and Eq.~\eqref{eq:SingleParticleSelfEnergyDiagram}, and the appearance of the effective exciton propagator $\mathcal{G}^\text{RPA}=\left(U^\text{eff}\right)^{-1} + \chi^\text{RPA}$ strengthens the analogy between this diagram and the usual single-particle self-energy diagram, which sums over internal virtual states. Indeed, it is the poles of this propagator at the collective mode energies which render the scattering rate contribution of this diagram non-zero. 
	
	\heading{Scattering below the gap} 
	Below the two-particle continuum, $\mathcal{V}$ is pole-free. Therefore, the only poles encountered in Eq.~\eqref{eq:ExcitonSelfEnergyDiagram} are those of $\chi^\mathrm{RPA}$ {which correspond to CMs}. We can thus simply results further, as we can retrace the same steps as in Appendix \ref{app:QuasiparticleSelfEnergy}.
	Combining Eqs.~\eqref{eq:chiSpectralPoleDecomposition} and \eqref{eq:SpectralADecomposition}, we have 
	\begin{equation}
	\chi^\text{RPA} \! \left(k,\omega\right) 
	= \frac{\mathcal{A}_1 \left(k\right)}{\omega^+ - \omega_k} \mathcal{R}_k 
	\mathrm{diag} (1,\, 0,\, 0,\, 0)
	\mathcal{R}_k^\d		\end{equation}
	This diagonal form allows us to factorize the scattering rate, which is given by the component $-\mathcal{A}_1\,\mathrm{Im}[\mathcal{R}_k^\d U^\mathrm{eff} \Sigma U^\mathrm{eff} \mathcal{R}_k]_{11}$ [cf. Eq.~\eqref{eq:ExcitonDecayRate}]. This yields
	%		\begin{widetext}
	\begin{align}
	\Gamma_k  & 		 = - \frac{1}{N} \!  \sum_{k'}  \mathrm{Im} \! \left[
	\frac{ \mathcal{A}_1 (k) \mathcal{A}_1 (k')}{\omega^+_k - \omega_{k'}} \right] 
	\tcbr \qquad  \times
	\Big\langle\Big|\mathcal{R}_k^\d U^\mathrm{eff} 
	\sum_{\lambda} \mathcal{V}^{\lambda} (k, k', \omega_k)  
	U^\text{eff}
	\mathcal{R}_{k'}\Big|_{11}^2 \Big\rangle \,,
	\end{align}
	%		\end{widetext} 
	where we have used the exchange hermiticity of $\mathcal{V}^\lambda$ below the gap. The imaginary part of the simple pole sieves out the density of collective modes and enforces the two on-shell scattering processes $k \rightarrow \pm k$,
	\begin{align}
	& -\mathrm{Im} \frac{1}{N} \sum_{k'} \frac{1}{\omega_k^+ - \omega_{k'}}
	\!\rightarrow \!\!
	\sum_{{k_0} = \pm k}  \int \!\! dk' \frac{\pi}{2} g_\mathrm{CM}(\omega_k)  \delta (k' - k_0).
	\end{align}
	$g_\mathrm{CM}\!\left(\omega_k\right) = 2 \times \frac{a}{2\pi} \Big|\frac{dk}{d\omega_k}\Big|$ is the density of CM states per site, the factor of two arising from the $\pm k$ degeneracy.
	{Assuming that $\mathcal{A}_1$ and $\mathcal{R}$ are parity-symmetric, we arrive at the scattering rate}
	\begin{align} \label{eq:CollectiveModeFermiGoldenRule1Dimension}
	\Gamma_k & 
	=\frac{1}{2} \pi g_\mathrm{CM} \!\left(\omega_k\right) \left[\mathcal{A}_1 \! \left(k\right)\right]^2  \tcbr \times \!\!\! \sum_{k'= \pm k} \!\!
	\Big\langle\Big|\mathcal{R}_k^\d U^\mathrm{eff} 
	\sum_{\lambda} \mathcal{V}^{\lambda} \left(k, k', \omega_k\right)  
	U^\text{eff}
	\mathcal{R}_{k}\Big|_{11}^2 \Big\rangle
	\end{align}
	Note this form is analogous to the FGR and the quasiparticle scattering rate \eqref{eq:SingleParticleDecayRateFull} we obtained previously. As before, we manually discard the forward-scattering contribution as it does not affect transport. 
	
	This result is reproduced in higher dimension, where kinematics make available additional scattering processes at arbitrary angles. Assuming anisotropy (at low energies), this generalizes to
	\begin{align} 
	\Gamma_k  & =  \pi g_\mathrm{CM} (\omega_k) \left[\mathcal{A}_1 (k)\right]^2 
	\tcbr \times \int \frac{d\Omega}{\Omega_D} 
	\Big\langle\Big|\mathcal{R}_k^\d U^\mathrm{eff} 
	\sum_{\lambda} \mathcal{V}^{\lambda} (k, k', \omega_k )  
	U^\text{eff}
	\mathcal{R}_{k}\Big|_{11}^2 \Big\rangle
	\end{align}
	where the integration is over solid angle $\Omega$ and $\Omega_D$ is the surface area of a unit $D$-sphere (we define $\Omega_1=2$). 
	The density of gapless CM states scales as $g_\mathrm{CM} \!\left(\omega\right) \sim v_g^{-1} k^{D-1} \sim \omega^{D-1}$.
	{As we now show,} at low energy $\mathcal{A}_1\left(k\right) \sim \omega_k^{-1}$, so that $\Gamma \sim \omega^{D-3} \times \text{(matrix element)}$. 
	
	%\end{widetext}
	\heading{Spectral weight}
	{That 		 $\mathcal{A}_1 \! \left(k\right) \sim \omega_k^{-1}$}
	is most easily seen in the particle--hole symmetric case: this symmetry protects the sectors $\mu,\nu=x,\,y,\,z$ and $\mu,\nu=0$ from mixing in $\chi^{0}$. We are then only interested in the former, as it hosts the collective modes. In this sector $U^\mathrm{eff} = (-U_0/2) \mathds{1}$, so that it commutes with $\chi^0$, and thereby $\chi^0,\,U^\mathrm{eff}$ and $\chi^\mathrm{RPA}$ share the same eigenbasis. However, $\chi^\mathrm{RPA}$ shares an eigenbasis with $\mathcal{A}_{\mu\nu}$ by construction. The latter has an eigenvalue $\mathcal{A}_1$ while the rest are zero. By Eq.~\eqref{eq:SpectralWeightFunctionASolution}, these are the imaginary parts of the eigenvalues of $\chi^\mathrm{RPA}$. Therefore, inspecting the RPA solution \eqref{eq:chiRPASolution} in this basis, the dominant eigenvalue of $\chi^\mathrm{RPA}$ satisfies
	\begin{align}
	\mathrm{Im} ~[ \lambda^\mathrm{RPA} (k, \omega)]
	& = \mathrm{Im} \left[ \left( \lambda^0 \!\left(k,\omega\right)\right)^{-1} + U_0/2 \right]^{-1}  
	\tcbr 
	= -\pi \mathcal{A}_1\!\left(k\right) \delta\left(\omega-\omega_k\right) \,,
	\end{align}
	where $\lambda^\mathrm{0,RPA}$ are eigenvalues of $\chi^\mathrm{0, RPA}$. 
	
	By Eq.~\eqref{eq:BareChi0Lehmann}$, \chi^0$ is Hermitian at energies below the two-particle continuum in the limit of vanishing $\eta$. Therefore, $\left(\lambda^0\right)^{-1}$ has an infinitesimal imaginary part, which in order to satisfy the relation above, must be positive. This yields
	\begin{align}
	\mathcal{A}_1 (k) \delta(\omega - \omega_k) 
	& = \delta \! \left(\mathrm{Re} \left( \lambda^0 (k,\omega)\right)^{-1} + U_0/2 \right)
	\tcbr = \left(\frac{2}{U_0}\right)^2 \left|\frac{\partial\omega}{\partial\lambda^0(k,\omega)}\right| \delta (\omega - \omega_k) \,,
	\end{align}
	and we find  that the CM frequency is  $\omega_k$ at which $\lambda^0 = -2/U_0$. 
	As the long-wavelength CMs are phase modes, $\lambda^0\left(k,\omega_k\right) = \chi^0_{yy} + \mathcal{O}\left(\omega_k^2\right)$. The retarded susceptibility satisfies $\chi^0\left(-\omega\right) = \chi^0\left(\omega\right)^*$, and as it is Hermitian and regular, we find $\chi^0_{yy}$ is real and even, so has derivative linear in $\omega$. \
	We obtain $\frac{\partial\lambda^0}{\partial\omega} \sim \omega + \mathcal{O}\left(\omega^2\right)$, thereby
	\begin{equation}
	\mathcal{A}_1\!\left(k\right) = \left(\frac{2}{U_0}\right)^2 \left|\frac{\partial\omega}{\partial\lambda^0}\right|_{\omega_k} \sim \omega_k^{-1}\,.
	\end{equation}
	The solution for $\mathcal{A}_1$ generalizes to the {particle--hole-asymmetric} case, and yields the same scaling.
	
	\heading{Channel $\lambda=0$ at $q\rightarrow0$}
	Channel $\lambda=0$ corresponds to a disordered electrostatic potential, and we expect that in this channel the collective modes would behave the most different compared to the single carriers, due to their neutral charge. Furthermore, we are interested in the long-wavelength dynamics of the collective modes. Note that $\mathcal{V}$ is only a function of $q$ and the center-of-mass momentum $\kappa = \left(k_1+k_2\right)/2$. The CMs scatter with momentum transfer $q=0$ and $q=-2k$, and for $k\rightarrow0$ the latter also tends to $q\rightarrow 0$. Therefore the scattering rate in this regime is dominated by the behaviour of $\mathcal{V}\left(q \rightarrow 0,\kappa,\omega\right)$.
	
	{We note the disorder screening (cf. Sec.~\ref{sec:Disorder}) mixes potential $\tilde{V}^{\lambda = 0}$ into the dressed potential $\tilde{\mathbb{V}}^{\lambda\neq0}$ via off-diagonal elements in $\chi_{\lambda0}\left(q, \omega=0\right)$. These elements will factor into $\mathcal{V}^{\lambda\neq0}$. However, as we mention in the main text, mixing between this channel and the rest vanishes at $q=0$ due to particle conservation. Therefore, for non-zero $q$ these elements scale as $\mathcal{O}\left(q^2\right)$. We now also show that $\mathcal{V}^0$ generates a $\mathcal{O}\left(q^2\right)$ contribution.}
	We expand for small $q$
	\begin{align}
	& {\mathcal{V}}^0_{\mu\nu}		  = - \frac{\tilde{\mathbb{V}}^0_q}{N} \!\sum_p \!\!\int \!\! \frac{d\omega '}{2 \pi i} \tr \Bigl\{\!
	\left[G_p^0(\omega')^2 + \frac{q}{2} \left[\partial_p G^0_p, \, G_p^0 \right]\!\!(\omega') \right] \tcbr \times
	\left[ \sigma^\mu G^0_{p+\kappa}(\omega' + \omega) \sigma^\nu + \sigma^\nu G^0_{p-\kappa}(\omega'-\omega) \sigma^\mu \right]
	\!\Bigr\} + \mathcal{O}(q^2)\,.
	\end{align}
	It can be shown that the leading term vanishes due to the contributions from the two diagrams in Fig.~\ref{fig:exciton_self_energy_vertex} canceling each other. As they corresponds to the disorder coupling to either the electron or hole, this is understood as a consequence of the exciton neutral charge. Therefore, $\mathcal{V}^0_{\mu\nu} = \mathcal{O}\left(q\right)$. For the term linear in $q$, $ \left[\frac{\partial}{\partial p}G^0_p, \, G_p^0 \right] \propto \sigma^y$, 
	so the linear term in $\mathcal{V}^0_{yy}$ leads to the trace 
	$\tr \left[\sigma^y \sigma^y G^0 \sigma^y  \right] = 0$, and therefore $\mathcal{V}_{yy}^0 = \mathcal{O}\left(q^2\right)$.
	
	At long wavelengths, the relevant components of $\mathcal{V}$ are 
	\begin{align}
	\big[\mathcal{R^\d V}^0 \mathcal{R} \big]_{11} &  = \mathcal{V}^0_{yy} + \mathcal{O}\left(\omega_k\right)\mathcal{V}^0_{y\nu} + \mathcal{O}\left(\omega_k^2\right) 
	\tcbr = \mathcal{O}\left(q^2\right) + \mathcal{O}\left(q \omega_k\right) + \mathcal{O}\left(\omega_k^2\right) = \mathcal{O}\left(k^2\right)\,.
	\end{align}
	{The quadratic dependence on $k$ is interpreted as follows: A constant electrostatic field would not affect the neutral excitons, so the coupling mechanism is between the exciton dipole moment and the gradient of the field, which is the second derivative of the disorder potential. This shows the relative factor of $k^4$ between the asymptotic scaling of channels $\lambda = z$ and $\lambda = 0$, even though they have the same $U(1)$ symmetry classification. }

	%%%%%%%%%%%%%%%%%%%%%%%%%%%%%%%%%%%%%%%%%%%%%%%%%%%%%%%%%%%%%%%%%
	%%%%
	%%%%			APPENDIX BREAK
	%%%%
	%%%%%%%%%%%%%%%%%%%%%%%%%%%%%%%%%%%%%%%%%%%%%%%%%%%%%%%%%%%%%%%%%

	\section{Collective Mode Disorder Scattering in Ginzburg--Landau Field Theory} \label{app:CollectiveModeFieldTheory}
	In this Appendix we will sketch how the scaling laws of the CM mean free path in the face of the various disorder channels could be explained in terms of an effective Ginzburg--Landau (GL) field theory \cite{AltlandSimons2010}. 
	
	\heading{Vector field theory}
	To capture the full phenomenology of the possible disorder coupling channels, we turn to a vector, rather than scalar, field theory. Motivated by the mapping onto spin-halves 
	$\vec{n}_i = \frac{1}{2} \Psi^\d_i \vec{\sigma} \Psi^{\phantom{\d}}_i$, we write down a Hamiltonian density for a fixed-length spin-$\frac{1}{2}$ field $\vec{n}\left(\vec{r}\right)$
	\begin{align}
	\mathcal{H} = \mu n_z  +  \frac{1}{2} &  \Lambda n^2_z + \frac{1}{2} \alpha_{xy} \left[\left(\del n_x \right)^2 + \left(\del n_y \right)^2\right] \tcbr + \frac{1}{2} \alpha_z \left(\del n_z \right)^2  + \vec{V}\!\left(x\right)\cdot \vec{n}\,.
	\end{align}
	The mapping between the mean fields of the main text and the spin field here is 
	$\lbrace n_x,\, n_y,\, n_z \rbrace \leftrightarrow \lbrace \mathrm{Re}~\braket{\phi},\, \mathrm{Im}~\braket{\phi},\, \braket{n_c} - \braket{n_v} \rbrace $. The first two terms define a ``Mexican hat'' potential wrapped on the $\mathbb{S}_2$ sphere, the two middle terms correspond to the stiffness of the fields, and the last term is the disorder potential. The original $U(1)$ symmetry is now manifest in the azimuthal symmetry of the clean portion of $\mathcal{H}$. Such an action could arise from a Hubbard--Stratonovich \cite{AltlandSimons2010} transformation of Eq.~\eqref{eq:Hubbard_Hamiltonian} via the direct ($c^\d_c c_c^{\phantom{\d}}$, etc.) and exchange ($c^\d_c c_v^{\phantom{\d}}$, etc.) channels.
	
	As the magnitude of the spin is fixed, a three-component description of the field is redundant, and only two are needed. The generalized coordinates are the azimuthal angle $\varphi$ and its canonically conjugate momentum $\Pi = n_z = S \cos \theta $, where $S$ is the spin magnitude and $\theta$ is the polar angle of the spin vector. The action of the theory is $\mathcal{S} = \int dt dx \mathcal{L}$ with Lagrangian density
	\begin{align}
	\mathcal{L} & 
	=  \Pi \dot{\varphi} -	\mu \Pi -\frac{\Lambda}{2}  \Pi^2  
	- \frac{ \alpha_{xy}}{2} \!\! \left[\!\frac{\left(\Pi \del \Pi \right)^2}{S^2 \!-\! \Pi^2} 
	\!+\! \left(S^2 \!-\! \Pi^2 \right)\left(\del \varphi\right)^2 \! \right]  \nonumber\\ & 
	- \frac{ \alpha_z}{2} \left(\del \Pi \right)^2 
	- V_z \Pi - \sqrt{S^2 \!-\! \Pi^2} \left(V_x \cos \varphi + V_y \sin \varphi \right)
	\,.
	\end{align}

	\heading{Stationary fields}
	Hamilton's equations read
	\begin{align}	
	&\dot\varphi   =  \mu + \Lambda \Pi  
	+ V_z - \frac{\Pi}{\sqrt{S^2 \!-\! \Pi^2}}\left(V_x \cos \varphi \!+\! V_y \sin \varphi \right)
	\tcbr 
	- \alpha_z \del^2 \Pi
	- \alpha_{xy} \left[\Pi \left(\del \varphi \right)^2  + \frac{\Pi^2 \del^2 \Pi}{S^2 \!-\! \Pi^2} 
	+ \frac{S^2 \Pi \left(\del \Pi \right)^2}{\left(S^2 \!-\! \Pi^2 \right)^2}\right],
	\\
	&\dot\Pi  = \del \! \left[ \alpha_{xy}  (S^2 \!-\! \Pi^2)\del \varphi \right] + \sqrt{S^2 \!-\! \Pi^2} (V_x \sin \varphi \!-\! V_y \cos \varphi ).
	\end{align}
	By setting the temporal derivates to zero, we can find the stationary configuration of the fields, which we will find to leading order in the disorder potential $V$. It is clear that in the absence of disorder, both fields are uniform, with $\Pi_0 = - \mu/\Lambda$, and $\varphi_0$ free in accordance with the unbroken $U(1)$ symmetry, and thus all derivates of the fields are already $\mathcal{O}\left(V\right)$. 
	
	In the presence of disorder the symmetry is generally spoiled. While weak disorder affects the magnitude of the order parameter negligibly, any infinitesimal strength will pin the phase $\varphi$. The phase which minimizes the free energy $H = \int dx \mathcal{H}$ is $\varphi_0$ which satisfies $\tan \varphi_0 = \bar{V_y} / \bar{V_x}$, where the over-bars represents the spatially-averaged value of the potentials. Therefore, we define the ``rotated'' potentials 
	\begin{equation}
	\tilde{V}_x = V_x \cos \varphi_0  + V_y \sin \varphi_0 , \quad 
	\tilde{V}_y = - V_x \sin \varphi_0 + V_y \cos \varphi_0 .
	\end{equation}
	Note that if $V_x$ and $V_y$ are correlated such that the disorder points in the same direction in all space, then $\tilde{V}_y$ vanishes identically. Therefore, $\tilde{V}_y$ is a potential corresponding to phase twisting. By coupling to $\varphi_0$, itself a random variable which is determined by the entire disorder realization,  $\tilde{V}_{x,y}$ are now spatially correlated even if the unrotated potentials are not. 
	
	Linearising $(\Pi,\, \varphi) =  (\Pi_1 (x),\, \varphi_1(x))  + \mathcal{O}(V^2)$, one obtains 
	\begin{align} \label{eq:FieldTheoryStationaryConfiguration}
	&   \alpha^0_\Pi  \del^2 \Pi_1 - \Lambda \Pi_1 = \mu + V_z - \frac{S^2\Pi_0}{({S^2 - \Pi_0^2})^{3/2}} \tilde{V}_x\,,
	\\
	&   \del^2 \varphi_1 = - \frac{1}{\alpha^0_\varphi} \sqrt{S^2 - \Pi_0^2} \tilde{V}_y = -\frac{1}{\alpha_{xy}} \frac{\tilde{V}_y}{\sqrt{S^2 - \Pi_0^2}}\,,
	\end{align}
	with $\alpha^0_\Pi = \alpha_z +\frac{\Pi^2_0}{S^2 - \Pi^2_0}  \alpha_{xy}$ and $\alpha^0_\varphi = \alpha_{xy}  \left(S^2 - \Pi_0^2\right) $.
	These equations are solved by usual Green's function methods. 
	This solution is morally equivalent to computing the linear response of the ground state of the field theory to the introduction of disorder, and mirrors the disorder screening discussed in Sec.~\ref{sec:Disorder}.

	\heading{Effective action}
	With the stationary configuration in hand, we can now expand the action in fluctuations about $\Pi_1, \varphi_1$. We substitute
	\begin{equation}
	\varphi = \varphi_1 + \delta\varphi\,, \qquad \Pi = \Pi_1 + \delta\Pi
	\end{equation}
	into the action, and retain only terms up to first power in $V$. By construction, the stationary configurations are those which eliminate terms linear in $\delta\Pi$ and $\delta\varphi$, and we can truncate at second order to obtain a quadratic effective action. For clarity we drop the $\delta$ signs, and obtain
	%\begin{widetext}
	\begin{align}
	\mathcal{L}_\mathrm{eff} 
	& = 
	\Pi \dot\varphi - 
	\frac{1}{2} \Lambda(x) \Pi^2 - \frac{1}{2}\alpha_\Pi(x) \left(\del \Pi \right)^2 
	\nonumber \\& 
	- \frac{1}{2} \alpha_\varphi (x) \left(\del\varphi\right)^2 
	%\hphantom{ = \int dt dx \Bigg\lbrace	\Pi \dot\varphi - \qquad}	
	+ \frac{1}{2} \tilde{V}_x(x)  \sqrt{S^2 \!-\! \Pi_0^2} \,\varphi^2 
	\nonumber \\& 
	+ \tilde{V}_y(x) \Pi  \left[\frac{\Pi_0}{\sqrt{S^2 \!-\! \Pi_0^2} } + 2 \alpha_{xy} \Pi_0 \left(\del\varphi_1\right) \del	\right] \varphi
	\end{align}
	%\end{widetext}
	where the spatially-modulated coefficients are linearised in $V$,
	\begin{align} \label{eq:FieldTheoryCoefficients}
	\Lambda(x) & = \Lambda -  \frac{2 \alpha_{xy} \Pi_0 S^2}{(S^2 - \Pi_0^2)^2}  \del^2 (\Pi_1  - \Pi_0)
	- \frac{S^2\tilde{V}_x}{({S^2 - \Pi_0^2})^{3/2}}  \,,   \\
	\alpha_\Pi (x) & =  \alpha^0_\Pi  + \frac{2 \alpha_{xy} \Pi_0 S^2}{\left(S^2 - \Pi_0^2\right)^2} \left(\Pi_1 - \Pi_0\right) \, ,\\
	\alpha_\varphi (x) & =  \alpha^0_\varphi - 2 \alpha_{xy} \Pi_0 \left(\Pi_1 - \Pi_0 \right) \,.
	\end{align}
	and represent terms which do not vanish in the absence of disorder. Note that $V_z$ no longer appears explicitly in the theory, but only through the modulation $\Pi_1 - \Pi_0$ in these coefficients (which also depends on $\tilde{V}_x$); this implies that this disorder channel becomes completely dressed by the symmetry-broken phase.

	\heading{Clean collective modes}
	From the effective action we can find the equations of motion, which are coupled and expressed
	\begin{equation}
	\dot{\vec{v}} = \hat{A}\vec{v} + \hat{B}\vec{v} 
	\end{equation}
	with the vector $\vec{v} = \left(\Pi_0\varphi, \Pi\right)^\mathrm{T}$, and $\hat{A}, \hat{B}$ generating the clean and disordered dynamics, respectively, and are 
	\begin{align}
	\hat{A} & = \begin{bmatrix}
	0 & \quad \Pi_0 \left(\Lambda^0 - \alpha^0_\Pi \del^2 \right) \\
	\frac{\alpha^0_\varphi}{\Pi_0} \del^2 & 0
	\end{bmatrix} \,, \\
	\hat{B} & = 
	\begin{bmatrix}
	\quad 0\qquad \Pi_0 \left(\delta\Lambda - \delta\alpha_\Pi \del^2 - \del\alpha_\Pi \del \right)  \\
	\frac{1}{\Pi_0} \left(\delta\alpha_\varphi \del^2 + \del \alpha_\varphi \del \right) \hfill 0\qquad
	\end{bmatrix} \tcbr + 
	\begin{bmatrix}
	0 & 0 \\
	1 & 0
	\end{bmatrix} \frac{\sqrt{S^2 - \Pi_0^2}}{\Pi_0} \tilde{V}_x   \tcbr-
	\begin{bmatrix}
	1 & 0 \\
	0 & 1
	\end{bmatrix} \left[ \frac{\Pi_0}{\sqrt{S^2-\Pi_0^2}} \tilde{V}_y - 2 \alpha_{xy} \Pi_0 \left(\del \varphi_1\right) \del \right]
	\,.
	\end{align}
	Transforming to Fourier space, we can find the eigenmodes of $\hat{A}$, which are 
	\begin{align}
	& \ket{\vec{v}^0_{k\pm} \! \left(t\right)}  = \varPhi_k \begin{pmatrix}
	1 \\ \mp i {\alpha_\varphi^0 k^2}/{\Pi_0 \omega_k}
	\end{pmatrix} e^{\mp i \omega_k t}\ket{k}\,, \tcbr
	\omega_k = \sqrt{\alpha^0_\varphi k^2 \left(\Lambda^0 + \alpha^0_\Pi k^2 \right)}\,,
	\qquad \left|\varPhi_k\right|^2 = \frac{1}{L^3} \frac{\hbar \omega_k }{\alpha_\varphi k^2 }\,.
	\end{align}
	Here the state $\ket{\vec{k}}$ is a plane wave $\exp{(i \vec{k}\cdot \vec{r})}$, and we fixed the normalization $\varPhi_k$ so that each mode carries one energy quantum of $\hbar \omega_k$. These are the Goldstone modes of the clean system. We can also glimpse that the mixing between the phase ($\varphi$) and amplitude ($\sim\Pi$) modes is proportional to $k^2/\omega_k \sim k$.
	
	%		\begin{widetext}
	\heading{Scattering theory}
	The disorder induces elastic scatting between the collective modes. We will evaluate this scattering rate via scattering theory \cite{Sakurai2017}. For simplicity we work in $D = 3$ dimensions. We assume an incident CM with wavenumber $k$. A true eigenstate of the system would be the superposition of the incident mode and all scattered modes of equal energy. Away from the region where the disorder is present, this energy can be evaluated and it coincides with $\omega_{k}$. Working within the Born approximation \cite{Sakurai2017}, the leading order solution for the true eigenmode is 
	\begin{equation}
	\ket{\vec{v}_k }= \ket{\vec{v}^0_k} + \hat{G}_k\hat{B}\ket{\vec{v}^0_k}\,,
	\end{equation}
	with the Green's operator $\hat{G}_k = (-i\omega_k \mathds{1} - \hat{A})^{-1} = (- \omega_k^2 \mathds{1} - \hat{A}^2 )^{-1} (-i\omega_k \mathds{1} + \hat{A})$.
	Executing $\hat{G}_k$, the correction to the state is
	\begin{align}
	\delta\vec{v}_k\!\left(\vec{r}\right) 
	& = \braket{\vec{r}|\delta\vec{v}_k} 
	%	= \int \frac{d^Dp}{\left(2\pi\right)^D} \braket{x|p}\! \braket{p | \hat{G}_k \hat{B}|\vec{v}^0_k}
	= \int \frac{d^D\vec{p}}{\left(2\pi\right)^D} \frac{e^{i \vec{p} \cdot \vec{r}}}{\omega_p^2 -\omega_k^2} \braket{\vec{p} | (\hat{A} -i\omega_k \mathds{1} )\hat{B}|\vec{v}^0_k}
	\tcbr = \int d\vec{r}' \frac{e^{i \left|\vec{k}\vphantom{'}\right| \left|\vec{r} - \vec{r}'\right|}}{\left|\vec{r}-\vec{r}'\right|} \vec{\rho}_k \left(\vec{r}'\right) \,
	\end{align}
	with the source term 
	\begin{align}
	\vec{\rho}_k (\vec{r}) 
	&\!=\!\!
	\begin{pmatrix}
	\rho_{\varphi, k} (\vec{r}) \\ \rho_{\Pi, k} (\vec{r})
	\end{pmatrix}
	\tcbr \!\!= \!\!
	\frac{\varPhi_k}{4\pi}  \frac{d(k^2)}{d(\omega_k^2)} 
	\braket{\vec{r} | (\hat{A} -i\omega_k \mathds{1} )\hat{B} \begin{pmatrix}
		1 \\ -i \frac{\alpha_\varphi^0 k^2}{\Pi_0 \omega_k}
		\end{pmatrix}|\vec{k}} \,.
	\end{align}
	
	The energy flux density is given by Noether's Theorem \cite{AltlandSimons2010}, and evaluates in the far field to
	\begin{align} \label{eq:FieldTheoryEnergyDensityDefinition}
	\vec{j}^\mathcal{E} & = \frac{\partial \mathcal L}{\partial \del \varphi} \dot 
	\varphi + \frac{\partial \mathcal L}{\partial \del \Pi} \dot \Pi 
	= - \alpha_\varphi  \dot \varphi \vec{\del}\varphi - \alpha_\Pi   \dot \Pi \vec{\del} \Pi
	\tcbr =  \alpha_\varphi \omega_k 
	\left[1 + \frac{\alpha_\Pi k^2}{\Lambda + \alpha_\Pi k^2}\right]\left[\left|\varPhi_k\right|^2 \vec{k}  +  \left|\tilde{\rho}_{\varphi,k}(\vec{k}')\right|^2\frac{ \vec{k}'}{r^2}\right]\,.
	\end{align}
	where $\vec{k}' = k \hat{r}$. Therefore $\tilde{\rho}_{\varphi,k}(\vec{k}')$, the Fourier transform of $\rho_{\varphi, k}$, gives the angular structure factor $f\left(\theta,\phi\right)$ of the scattered wave, yielding the differential scattering cross-section 
	\begin{align}
	\frac{d\sigma}{d\Omega} & = \frac{j_\mathrm{out}}{j_\mathrm{in}} = \left|\rho_{\varphi,k} ( \vec{k}' ) / \varPhi_k \right|^2 
	\tcbr =  \left|\frac{1}{4\pi}\frac{d(k^2)}{d(\omega_k^2)} 
	\braket{\vec{k}' | \left(1 \ 0\right) \left[\hat{A} -i\omega_k \mathds{1} \right]\hat{B}\begin{pmatrix}
		1 \\ -i \frac{\alpha_\varphi^0 k^2}{\Pi_0 \omega_k}
		\end{pmatrix}|\vec{k}}\right|^2 \!.
	\end{align}
	Finally, we can obtain the scattering rate from the incoming particle density and the integrated cross-section
	\begin{equation}
	\Gamma_k 
	= \frac{1}{\hbar \omega_k} j^{\mathcal{E}}_\mathrm{in} \sigma 
	= \left[1 + \frac{\alpha_\Pi k^2}{\Lambda + \alpha_\Pi k^2}\right] \frac{1}{L^3} \frac{\omega_k }{k }\times \sigma\,.
	\end{equation}

	\revised{\heading{Long-wavelength limit}}
	We can find the asymptotic scaling of $\Gamma_k$ at very small momentum $k$. In this limit the dispersion is approximately linear with $\omega_k = \sqrt{\Lambda^0 \alpha^0_\varphi} k = c k$. Therefore $\Gamma_k \sim c \sigma L^{-3} $ and the differential cross-section evaluates to 
	\begin{align}
	\frac{d\sigma}{d\Omega} &  = \frac{1}{4 \pi c^2}\Bigg|
	- (\vec{k}'\cdot \vec{k})\big[\alpha_\varphi^0 k^2 \braket{k'|\delta\alpha_\Pi|k'} 
	\tcbr  +
	\left(\Lambda^0 + \alpha^0_\Pi k^2 \right) \braket{k'|\delta \alpha_\varphi|k}   \big] 
	- k^2 \alpha_\varphi^0 \braket{k'|\delta\Lambda|k}
	\tcbr +
	\sqrt{S^2 - \Pi_0^2} \left(\Lambda^0 + \alpha^0_\Pi k^2 \right) \braket{k'|\tilde{V}_x  |k} 
	\tcbr -2i\omega_k 
	\left[1 + 2  \frac{(\vec{k}'-\vec{k}) \cdot \vec{k} }{(\vec{k}'-\vec{k})^2}\right]\frac{\Pi_0}{\sqrt{S^2 - \Pi_0^2}} \braket{k'|\tilde{V}_y|k} 
	\Bigg|^2\,.
	\end{align}
	Here the first two terms correspond to the dressing of the disorder in the stationary configuration, while the last two correspond to the direct effect of the disorder channels.
	It is immediately seen that the contribution of channel $\tilde{V}_y$ vanishes, as the the scattering is elastic and $\left|k'\right| = \left|k\right|$. Channel $\tilde{V}_x$ contributes directly by $\braket{\vec{k}'|\tilde{V}_x|\vec{k}} = \tilde{V}_x\!\left(\vec{q}\right)$ with $\vec{q} = \vec{k}'-\vec{k}$. 
	We note that all coefficients are modulated by $\left(\Pi_1 - \Pi_0\right)$, and by the stationary configuration condition \eqref{eq:FieldTheoryStationaryConfiguration}, 
	$
	\braket{\vec{k}'|\left(\Pi_1 - \Pi_0\right)|\vec{k}} 
	\sim \braket{\vec{k}'|V_{x,z}|\vec{k}}
	$
	so that these terms contribute
	$
	\sim k^2 V_z\left(\vec{q}\right) + k^2 \tilde{V_x}\left(\vec{q}\right) + \mathcal{O}\left(k^4\right)
	$.
	
	Collecting contributions, we find 
	\begin{align}
	\Gamma_k & 
	\sim \left|\tilde{V}_x \left(\vec{q}\right) +  k^2 V_z\left(\vec{q}\right) + k^2 \tilde{V_x}\left(\vec{q}\right) + \mathcal{O}\left(k^4\right)  \right|^2 \tcbr
	\sim \left|\tilde{V}_x \left(\vec{q}\right)\right|^2  
	+ k^2  \mathrm{Re}\left[\tilde{V}_x \left(\vec{q}\right) V^{*}_z \left(\vec{q}\right)\right] 
	%+ k^4 \left|\tilde{V}_x \left(q\right)\right|^2 
	+ k^4\left|V_z  \left(\vec{q}\right)\right|^2  \,.
	\end{align}
	Recognizing that the scattering rate is proportional to the CM density of states $g_\mathrm{CM}\sim k^{D-1}$, which is quadratic in three dimensions, we may then generalize to arbitrary dimension $D$. Assuming the different disorder channels are uncorrelated, the cross term vanishes and we obtain 
	\begin{equation} \label{eq:FieldTheoryScatteringRateScaling}
	\Gamma_k 
	\sim 
	k^{D-3} \left|\tilde{V}_x \left(\vec{q}\right)\right|^2  
	+ k^{D+1}\left|V_z  \left(\vec{q}\right)\right|^2  \,.
	\end{equation}
	
	%		\end{widetext}
	
	%%%%%%%%%%%%%%%%%%%%%%%%%%%%%%%%%%%%%%%%%%%%%%%%%%%%%%%%%%%%%%%%%
	%%%%
	%%%%			APPENDIX BREAK
	%%%%
	%%%%%%%%%%%%%%%%%%%%%%%%%%%%%%%%%%%%%%%%%%%%%%%%%%%%%%%%%%%%%%%%%

	\section{Collective Mode Thermal Conductivity} \label{app:CollectiveModeThermodynamics}
	In this Appendix we evaluate the CM contribution to thermodynamic quantities such as specific heat and thermal conductivity as a function of temperature. 
	{As massless bosonic excitations, their contribution would be similar to that of ordinary phonons, and we will model it similarly to the Debye model of phonons \cite{Kittel}. We assume that the temperature is sufficiently low compared to the EI transition temperature  so that the dispersion of the CMs is well-characterised by its zero temperature limit, and that the CMs can be viewed as a low density gas of composite bosons \cite{MoskalenkoSnoke2000}. Since the EI ordering temperature can be large, up to room temperature in materials such as \TNS{}, these conditions can be readily met over a wide range of temperatures.}	
	
	We will work in arbitrary dimension $D$. The CM contribution to the specific heat is 
	\begin{align}
	\frac{C_V}{N k_B} 
	& = - \frac{\beta^2}{N} \sum_k \hbar \omega_k \frac{\partial}{\partial \beta}  f_\mathrm{BE}\left(\hbar\omega_k, T\right) \tcbr 
	= \frac{1}{N}\! \sum_k \! \left[\frac{\beta \hbar \omega_k}{e^{\beta \hbar \omega_k} - 1}\right]^2 
	\!\!\rightarrow \!\!
	\int d\omega g_\mathrm{CM} \! \left(\omega\right) \left[\frac{\beta \hbar \omega}{e^{\beta \hbar \omega} - 1 }\right]^2
	\end{align}
	where $f_\mathrm{BE}$ is the Bose-Einstein distribution and $\beta$ is the inverse temperature. The squared bracket then represents the contribution of each mode to the specific heat. 
	From this we can derive the thermal conductivity \cite{Kittel}
	\begin{align}
	\frac{\kappa}{N} &
	= \frac{1}{N D} \sum_k v_g (k) l_\mathrm{mfp}(k)  C_V(k) \tcbr
	= \frac{k_B}{N D}  \sum_k \frac{v^2_g  (k)}{\Gamma (k)}  \left[\frac{\beta \hbar \omega_k}{e^{\beta \hbar \omega_k} -1 }\right]^2 
	\tcbr \rightarrow \frac{k_B}{D} \int d\omega g_\mathrm{CM} (\omega) \frac{v^2_g (\omega)}{\Gamma(\omega)}  \left[\frac{\beta \hbar \omega}{e^{\beta \hbar \omega} -1 }\right]^2 .
	\end{align}
	We now substitute into these expressions our Debye model assumption of dispersionless CMs with $\omega_k = v_g \left|k\right|$ on an isotropic spherical BZ. The CM DoS is then normalized $g_\mathrm{CM}\left(\omega\right) = \frac{D}{\omega} \left(\frac{\hbar \omega}{k_B \Theta_\mathrm{CM}}\right)^D$, {with $\Theta_\mathrm{CM}$ the CM Debye temperature, corresponding to the CM band width. Our focus is then on $T\ll\Theta_\mathrm{CM}$.} 
	Plugging in these expressions, we find the usual Debye result for the specific heat
	\begin{equation}
	\frac{C}{N k_B} = D \left(\frac{T}{\Theta_\mathrm{CM}}\right)^D \int_0^{\Theta_\mathrm{CM}/T} \frac{x^{D+1}dx}{\left(e^x-1\right)^2}
	\, \sim \, T^D\,.
	\end{equation}
	
	To evaluate the thermal conductivity, we substitute the scaling relations,  $\Gamma \sim \omega^{D-\nu}$ where $\nu$ depends on the disorder coupling channel. It takes the values $\nu = 3$ and $\nu = -1$ for symmetry-violating and symmetry-preserving disorder, respectively [cf. Appendix \ref{app:CollectiveModeFieldTheory} and Eq.~\eqref{eq:FieldTheoryScatteringRateScaling}]. We argue in Sec.~\ref{sec:Discussion} that for charge-coupled disorder, exciton charge neutrality further increases the scaling to $\nu = -5$. This is supported by numerics in one dimension, and a long-wavelength expansion in Appendix \ref{app:ExcitonSelfEnergyDiagrammatics}.
	
	As the scattering rate induced by symmetry-preserving disorder vanishes in the long-wavelength limit, the mean free path is truncated by the system size. We thus construct the scattering rate 		\begin{equation}
	\Gamma = \frac{K_0}{\hbar J} \left(\frac{\hbar \omega}{\Delta}\right)^{D-\nu} F_\nu + \frac{v_g}{L}\,.
	\end{equation}
	where $\Delta$ is the gap and $L$ is the system linear size. $F_\nu$ is a dimensionless constant of order 1 in channels $\lambda = x,z$, while for channel $\lambda = 0$ it is a function of the distance from particle--hole symmetry $\left(J_c - J_v\right)$ which vanishes at the symmetry point.
	
	For the symmetry-violating channel, disorder dominates the scattering, and we can take $L\rightarrow \infty$. The thermal conductivity becomes 
	\begin{align}
	\frac{\kappa}{N}  &		= k_B v_g^2   \left(\frac{\Delta}{k_B \Theta_\mathrm{CM}}\right)^D \left(\frac{k_B T}{\Delta} \right)^3 \frac{\hbar J }{K_0 F_x} 
	\int_0^{\frac{\Theta_\mathrm{CM}}{T}}\!\! \frac{x^{4}dx}{\left(e^x \!-\! 1\right)^2} % \\
	\tcbr \propto  k_B v_g a \, \frac{J^2}{K_0} \left( \frac{T}{\Theta_\mathrm{CM}} \right)^3 \sim T^3 / K_0 \,.
	\end{align}
	
	By contrast, with symmetry-preserving disorder, we must distinguish between modes whose free path is limited by disorder and modes limited by the system size. The thermal conductivity is then written
	\begin{align}
	\frac{\kappa}{N}  		 \approx   k_B v_g L  & \left(\frac{T}{\Theta_\mathrm{CM}}\right)^D \Bigg[ 
	\int_0^{\frac{\Theta_\mathrm{XO}}{T}}  \frac{x^{D+1}dx}{\left(e^x - 1\right)^2}  \tcbr
	+ \left(\frac{\Theta_\mathrm{XO}}{T}\right)^{D-\nu} \int_{\frac{\Theta_\mathrm{XO}}{T}}^{\frac{\Theta_\mathrm{CM}}{T}} \frac{x^{\nu+1}dx}{\left(e^x - 1\right)^2} 
	\Bigg]
	\end{align}
	where
	\begin{equation*}
	k_B \Theta_\mathrm{XO} = \Delta \left[\frac{\hbar v_g J}{F_\nu K_0 L}\right]^{\frac{1}{D-\nu}} \sim k_B \Theta_\mathrm{CM} \left[\frac{1}{F_\nu} \frac{J^2}{K_0} \frac{a}{L}\right]^{\frac{1}{D-\nu}}
	\end{equation*}
	is the crossover temperature between the regimes dominated by system size and disorder scattering, represented by the first and second terms, respectively. 
	
	We evaluate the integrals asymptotically and find the scaling relations
	\begin{equation}
	\frac{\kappa}{N} \approx 
	\begin{cases}
	\hphantom{\frac{1}{\left|v\right|}} k_B	 v_g L \left(\frac{T}{\Theta_\mathrm{CM}}\right)^D  \sim L T^D, & T < \Theta_\mathrm{XO} \\		\frac{1}{\left|\nu\right|} k_B v_g L  \left(\frac{\Theta_\mathrm{XO}}{\Theta_\mathrm{CM}}\right)^D    \sim K_0^{-\frac{D}{D-\nu}} L_{\vphantom{0}}^{-\frac{\nu}{D-\nu}}
	, & \Theta_\mathrm{XO} < T
	\end{cases}
	\end{equation}
	Note that for the symmetry-preserving channels $\nu < 0 $. Here we assumed that the disorder scattering, though weak, is strong enough such that $\Theta_\mathrm{XO} < \Theta_\mathrm{CM}$, so that the high-temperature regime above is still realized in the condensate at temperatures well below $\Theta_\mathrm{CM}$. This result implies that heat is transported only by modes under energy $\Theta_\mathrm{XO}$, as the modes above  are comparatively strongly scattered; therefore, above the crossover temperature these modes' specific heat is saturated and so is the thermal conductivity.

	This treatment ignores other scattering mechanisms which would become dominant at higher temperatures. As an example, we consider the effect of CM--phonon scattering in a system which is free of intrinsic symmetry-violating disorder. We assume that the CMs are coupled to an optical phonon branch with Einstein frequency $\omega_\mathrm{E}$. The thermal phonons are incoherent so that we consider their instantaneous configuration as a static disorder potential in which the CMs propagate. This picture holds if the phonons are macroscopically excited, i.e. at $T \ge \Theta_\mathrm{E} = \hbar\omega_\mathrm{E} / k_B$.  In general the CM--phonon coupling may violate the condensate symmetry, and it follows that the thermal conductivity scales as $\kappa \sim T^3 / f_{\mathrm{BE}} \! \left(\omega_0 ; T\right)$. Above the Einstein temperature $\kappa \sim T^2$, while below it the scattering by thermal phonons is exponentially suppressed, and the mean free path is dominated by symmetry-respecting disorder, the system size, etc., and $\kappa$ reflects the CM specific heat dependence. This would then manifest as a peak in the thermal conductivity of the CMs in the crossover between the two regimes, as sketched schematically in Fig.~\ref{fig:thermal_conductivity}.

	We observe that this is a result of the separation of energy scales between the CM and phonons, as there exists an intermediate temperature range where the CM have an appreciable thermal population and yet the phonons do not. This upturn in the thermal conductivity can be interpreted as a general feature resulting from the crossing of fast, low-lying modes with other massive degrees of freedom. We note that such an increase in thermal conductivity with decreasing temperature was measured in the EI candidate \ce{TmSe_{0.45}Te_{0.55}} \cite{Wachter2004}, where it was attributed to EI superfluidity; here we have shown a different mechanism which would produce this signature.

	%		\end{document}

\end{document}